\documentclass{MITcsail}

\title{Model-free quantification of completeness, uncertainties, and outliers in atomistic machine learning using information theory}

\author
{Daniel Schwalbe-Koda~$^{1, 2}$\footnote{E-mail: dskoda@ucla.edu},  Sebastien Hamel~$^{1}$, Babak Sadigh~$^{1}$, Fei Zhou~$^{1}$, Vincenzo Lordi~$^{1}$\footnote{E-mail: lordi2@llnl.gov}\\
\vspace{1em} 
\normalfont{\small $^{1}$Lawrence Livermore National Laboratory, Livermore, CA 94550, United States}\\
\normalfont{\small $^{2}$University of California, Los Angeles, Los Angeles, CA 90095, United States}
}

\usepackage[resetlabels]{multibib}
\newcites{Supp}{Supplementary References}
\newcites{Main}{References}

\usepackage{booktabs}
\usepackage{dsk}

\begin{document}

\maketitle
\thispagestyle{firstpagestyle} 

\begin{abstract}
An accurate description of information is relevant for a range of problems in atomistic machine learning (ML), such as crafting training sets, performing uncertainty quantification (UQ), or extracting physical insights from large datasets.
However, atomistic ML often relies on unsupervised learning or model predictions to analyze information contents from simulation or training data.
Here, we introduce a theoretical framework that provides a rigorous, model-free tool to quantify information contents in atomistic simulations.
We demonstrate that the information entropy of a distribution of atom-centered environments explains known heuristics in ML potential developments, from training set sizes to dataset optimality.
Using this tool, we propose a model-free UQ method that reliably predicts epistemic uncertainty and detects out-of-distribution samples, including rare events in systems such as nucleation.
This method provides a general tool for data-driven atomistic modeling and combines efforts in ML, simulations, and physical explainability.
\end{abstract}

\section{Introduction}

\begin{figure*}[!h]
    \centering
    \includegraphics[width=0.8\linewidth]{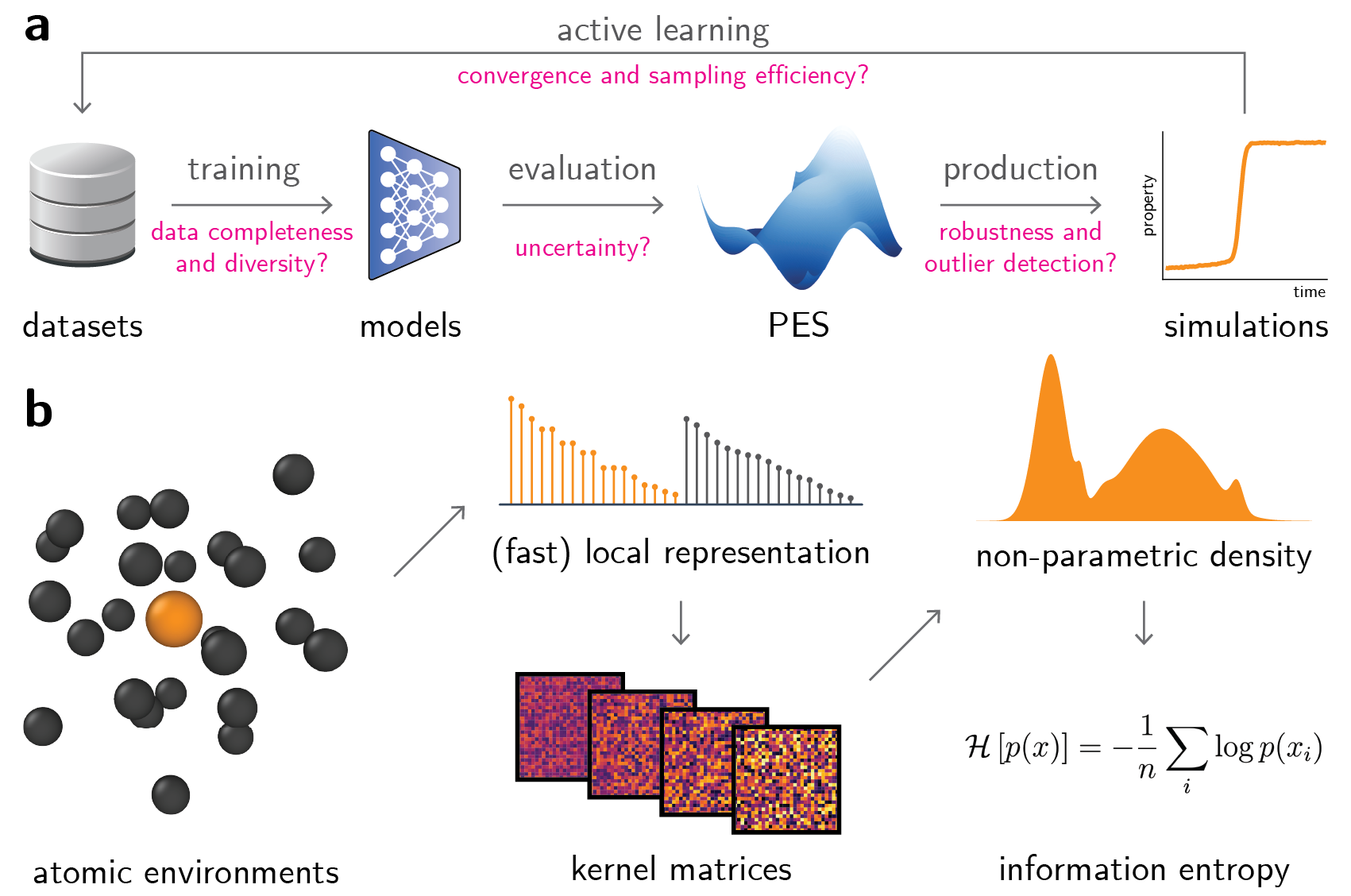}
    \caption{
    \takeaway{Overview of the method.}
    \textbf{a}, Typical workflow in MLIPs for training, evaluating, and retraining models that predict potential energy surfaces (PES). Challenges in the process are highlighted in magenta.
    \textbf{b}, Overview of our QUESTS method, which computes the information entropy of a non-parametric descriptor distribution.
    }
    \label{fig:01-overview}
\end{figure*}

Quantifying information contents in datasets is an essential task in many fields of science.
Information theory, initially proposed in the context of communication theory, \citeMain{shannon1948mathematical} enabled a rigorous treatment of data, errors, and information in fields beyond its own, such as statistical thermodynamics,\citeMain{jaynes1957information} biophysics,\citeMain{rashevsky1955life} or deep learning.\citeMain{shwartz2017opening}
Particularly in the materials and chemical sciences, the known parallels between thermodynamic entropy and information theory\citeMain{shannon1948mathematical} have long inspired quantitative descriptions of information contents in atomistic data.
Simulation outcomes have been extensively related to thermodynamic entropy,\citeMain{wallace1987Correlation,baranyai1989Direct,morris1995Calculatinga,vansiclen1997Information,vink2002Configurational,killian2007Extraction,huang2022Vibrationala} often with an explicit definition of the degrees of freedom for the systems.
Nevertheless, the need for an information metric in atomistic data goes beyond that from statistical thermodynamics, especially within modern computational materials science.
For instance, machine learning interatomic potentials (MLIPs) have showed great promise in bypassing density functional theory (DFT) in atomistic simulations, \citeMain{Behler2007,Bartok2010,Behler2015,thompson2015spectral,Chmiela2017,zhang2018deep,Mueller2020Machine,Manzhos2020a,Unke2021a,Batzner2022,Batatia2022} but their development requires careful dataset construction, reliable uncertainty quantification (UQ) strategies, and statistical tools to assess the reliability of production simulations to reflect the chosen (e.g., DFT) ground truth (Fig. \ref{fig:01-overview}a).
At some level, these tasks require assessing how much information is present in the input or output data.
These requirements are particularly critical in the case of neural network (NN)-based MLIPs, whose costly training processes and unreliable extrapolation performances require efficient training techniques and improved reliability in predictions. \citeMain{schwalbekoda2021differentiable,Fu2022,vita2023losslands}
Furthermore, detection of rare events, order parameters, and accessible phase spaces is an important task within atomistic simulations, but often requires hand-crafted descriptors.
Thus, understanding information contents within atomistic data and rigorously quantifying them is essential for improving training efficiency, robustness, and interpretability of machine learning (ML)-driven simulations.

In this work, we propose a method to quantify information contents in atomistic simulation data.
Building on the concepts of information theory, we show that the information entropy from a distribution of atom-centered representations explains trends in MLIP errors, rationalizes dataset analysis/compression, provides a robust UQ estimate for ML-driven simulations, and can be used for outlier detection in a number of  applications.
In contrast with other works connecting information entropy and atomistic data, this work proposes a framework to match the assumptions of information theory, thus inheriting its useful properties in data compressibility, efficiency, error detection, and others.
Using this formalism, we show how information contents can be used to extract physical insights from atomistic datasets, including correlations in error metrics for datasets or explaining outliers in trajectories, while avoiding the need for a trained model.
This work provides a new tool for atomistic simulations, MLIP development, and UQ for computational materials science, and can be extended to enable faster and more accurate materials modeling beyond predictions of potential energy surfaces (PES).

\section{Results}

\subsection{Formulation of an atomistic information entropy}
\label{sec:repr}

To approximate a one-to-one mapping between atomistic environments and the probability distributions from which the data is sampled, we propose a descriptor for atomic environments inspired by recent studies in continuous and bijective representations of crystalline structures\citeMain{Widdowson2022ResolvingData} and similar to the DeepMD descriptors.\citeMain{zhang2018deep}
Given their success in a range of applications,\citeMain{zhang2018deep,schwalbekoda2023inorganic} the representation offers a rich metric space without sacrificing its computational efficiency, allowing us to perform non-parametric estimates of data distributions even in large datasets ($> 10^6$ atoms).
Despite its simplicity, it also mitigates discontinuities in the representation latent space within controlled hyperparameter settings (see \supptext).
To obtain the descriptor, we begin by sorting the distances from a central atom $i$ to its $k$-nearest neighbors (within periodic boundary conditions, if appropriate) and obtain a vector $\X_i^{(1)}$ with length $k$,

\begin{equation}\label{eq:x1-short}
    \X_i^{(1)} = \begin{bmatrix}
    \frac{w(r_{i1})}{r_{i1}} & \ldots & \frac{w(r_{ik})}{r_{ik}}
    \end{bmatrix}^T
,~
    r_{ij} \leq r_{i(j + 1)},
\end{equation}

\noindent with $1 \leq j \leq k$ due to the $k$-nearest neighbors approach and $w$ a smooth cutoff function given by

\begin{align}\label{eq:cutoff}
    w(r) = \begin{cases}
        \left[
            1 - \left(\frac{r}{r_c}\right)^2
        \right]^2 ,& 0 \leq r \leq r_c, \\
        0, & r > r_c.
    \end{cases}
\end{align}

\noindent However, the radial distances alone do not capture bond angles and cannot be used to fully reconstruct the environment, so we construct a second vector to augment $\X_i^{(1)}$ that aggregates distances from neighboring atoms, inspired by the Weisfeiler-Lehman isomorphism test and analogous to message-passing schemes in graph neural networks (see \methods\ and Fig. \ref{fig:si:01-descriptors}):

\begin{equation}\label{eq:x2-short}
    X_{in}^{(2)} =
    \left\langle
    \frac{\sqrt{w(r_{ij}) w(r_{il})}}{r_{jl}}
    \right\rangle_n
    ,~ j, l \in \mathcal{N}(i),~ X_{in} \geq X_{i(n + 1)},
\end{equation}

\noindent where $j$ and $l$ are atoms in the neighborhood $\mathcal{N}$ of atom $i$, $\langle . \rangle_n$ represents the average of the $n$-th elements of the sequence (details in the \supptext), and $1 \leq n \leq k - 1$ due to the number of  $k$-nearest neighbors pairs.
The final descriptor $\X_i$ is obtained by concatenating $\X_i^{(1)}$ and $\X_i^{(2)}$.
As this representation requires only the computation of a neighbor list, it can be easily parallelized and scaled to large systems.

To quantify the information entropy from a distribution of feature vectors $\Xset$, we start from the definition of the Shannon entropy $\Info$,\citeMain{shannon1948mathematical}

\begin{equation}\label{eq:entropy-shannon}
   \Info\left[p(x)\right]  = - \int p(x) \log p(x) dx,
\end{equation}

\noindent where $\log$ is the natural logarithm, implying that $\Info$ is measured in units of nats.
Given a kernel $K_h$ with bandwidth $h$, we can perform a kernel density estimate (KDE) of a distribution of $n$ atomic environments $\X_i$ to obtain a non-parametric estimate of the information entropy of $p(x)$,\citeMain{Beirlant1997Nonparametric}

\begin{equation}\label{eq:entropy-main}
    \HX = - \frac{1}{n} \sum_{i=1}^n
        \log \left[
            \frac{1}{n}
            \sum_{j=1}^n K_h (\X_i, \X_j)
        \right],
\end{equation}

\noindent which corresponds to a discrete version of the original information entropy in Eq. \eqref{eq:entropy-shannon} (see derivation in the \supptext).
If the kernel is defined in the space $K_h: \mathbb{R}^N \times \mathbb{R}^N \rightarrow [0, 1]$, then the entropy from Eq. \eqref{eq:entropy-main} recovers useful properties from information theory such as well-defined bounds ($0 \leq \Info \leq \log n$) and quantifies the absolute amount of information in a dataset $\Xset$.
This agreement with information-theoretical definitions contrasts with other relative metrics of entropy in atomistic datasets such as the ones from Perez \textit{et al.}\citeMain{Karabin2020EntropyMaximization} or Oganov and Valle,\citeMain{oganov2009Howa} which can be ill-defined in the presence of identical feature vectors or structural outliers.
In our definition, $\Info = \log n$ implies $K_h(\X_i, \X_j) = \delta_{ij}$, which is the case when all points are dissimilar from each other.
$\Info = 0$, on the other hand, implies $K_h(\X_i, \X_j) = 1,\ \forall i, j$, which represents a degenerate dataset with all points equivalent to each other.
We discuss the importance of these and other useful properties of the information entropy for atomistic simulations in detail at the \supptext.

In this work, we choose $K_h$ to be a Gaussian kernel,

\begin{equation}\label{eq:kernel-gaussian}
    K_h (\X_i, \X_j) = \exp \left(
        \frac{-\norm{\X_i - \X_j}^2}{2h^2}
    \right),
\end{equation}

\noindent where the bandwidth $h$ is selected to rescale the metric space of $\X$ according to the distance between two FCC environments with a 1\% strain (\supptext, Sec. \ref{sec:si:bw}).
Nevertheless, as the choice of kernel is known to influence the estimated distribution, the bandwidth (and associated entropy) may vary according to the kernel.
Within this work, the bandwidth was kept constant, and was found to be reasonably adequate for all the ML-related tasks that adopt this Gaussian kernel.

To quantify the contribution of a data point $\Y$ to the total entropy of the system, we define the differential entropy $\dH$ as

\begin{equation}\label{eq:dH}
    \dH(\Y | \set{\X_i} ) = - \log \left[
        \sum_{i=1}^n K_h (\Y, \X_i)
    \right],
\end{equation}

\noindent where $\dH$ is defined with respect to a reference set $\Xset$ and can assume any real value.
The measure $\dH$ intuitively represents how much ``surprise'' there is in a new point $\Y$ given the existing observations $\set{\X_i}$, and will be shown to enable uncertainty quantification, outlier detection, and other important results.

An overview of this method, named Quick Uncertainty and Entropy from STructural Similarity (QUESTS), is shown in Fig. \ref{fig:01-overview}b, and a range of examples and visualizations demonstrating the intuition behind the method are provided in the \supptext\ (Figs. \ref{fig:si:01-x-scaling}--\ref{fig:si:01-lindemann}). The code is available at \url{https://github.com/dskoda/quests}.

\subsection{Information-theoretical dataset analysis for machine learning potentials}

Most recent MLIPs predict potential energy surfaces from fixed or learned atom-centered representations, similar to the strategy adopted in this work.
Despite the wide usage of these models, constructing training datasets for these potentials is still a challenge.\citeMain{Smith2021,bernstein2019novo,MontesdeOcaZapiain2022,qi2023robust}
Works such as the ones from Perez \textit{et al.} proposed quantifying entropy as a way to build diverse atomistic datasets,\citeMain{Karabin2020EntropyMaximization,MontesdeOcaZapiain2022} but their approximation to entropy in the descriptor space prevents recovering true values of information from datasets, as defined by information theory.
Furthermore, while using large amounts of data to train models can enhance the generalization power of NNIPs,\citeMain{chen2022universal,merchant2023scaling,batatia2023foundation} na\"ively generating large, redundant datasets can be a counterproductive strategy, and training costs can be reduced by creating smaller optimal datasets that still achieve similar or even better results.\citeMain{qi2023robust}
Since generating training data generally involves computationally expensive ground truth calculations and training a model on a larger dataset leads to increased training cost and higher training complexity, there is significant advantage in understanding how to simultaneously minimize dataset size, maximize their coverage in the configuration space, and still maintain the accuracy of the MLIP trained on the full dataset.

\subsubsection{Relating information contents to learning curves in molecular datasets}

\begin{figure*}[!h]
    \centering
    \includegraphics[width=\linewidth]{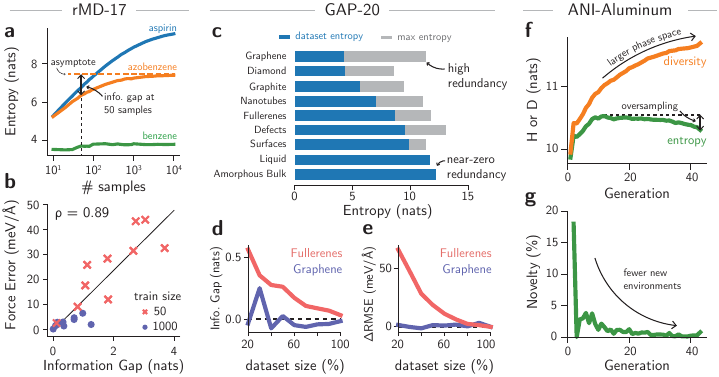}
    \caption{
    \takeaway{Information entropy measures dataset completeness, compressibility, and sample efficiency in MLIPs.}
    \textbf{a}, Information entropy for three example molecules from the rMD17 dataset as a function of the dataset size.
    Simpler molecules exhibit lower entropy and converge faster, while more diverse molecules require more samples to converge.
    \textbf{b}, correlation between the error in predicted forces and the information gap for all molecules in the rMD17.
    The errors were obtained from the original reference for MACE.
    A circle indicates errors when 1000 samples are used to train the models, and crosses are errors when only 50 samples are used to train the models.
    $\rho$ is the Pearson's correlation coefficient.
    \textbf{c}, information entropy (blue bars) of selected subsets of the carbon GAP-20 dataset.
    The maximum entropy is given by $\log n$ (gray bars), where $n$ is the number of atomic environments.
    The results are sorted by ascending dataset entropy.
    \textbf{d}, information gap obtained by compressing the ``Fullerenes'' and ``Graphene'' subsets of GAP-20 by up to 20\% of their original sizes.
    While the information gap of ``Graphene'' remains close to zero, the one from ``Fullerenes'' monotonically increases as the dataset size decreases.
    \textbf{e}, test errors relative to the errors obtained when a MACE model is trained on the full subset of GAP-20.
    The results show that the ``Graphene'' subset can be compressed by up to 20\% of its size without loss of performance, whereas this is not the case for the ``Fullerenes'' subset.
    \textbf{f}, information entropy and diversity for the ANI-Al dataset computed for each generation of active learning.
    Oversampling of certain phases leads to a total reduction of entropy, as demonstrated by \textbf{g}, showing decreasing novelty in the samples.
    In this approach, novelty is the fraction of environments showing
    $\dH > 0$ when the dataset of all previous generations
    are taken as reference.
    Nevertheless, the diversity of the dataset continues to increase.
    }
    \label{fig:02-dset}
\end{figure*}

Borrowing from a fundamental concept in information theory, \takeaway{we hypothesize that the information entropy of atomistic datasets relates to the limit of their (lossless) compression} and can thus explain results from learning curves in MLIPs.
The theoretical results from information theory already guarantee the compression limits that can be applied to any data,\citeMain{shannon1948mathematical} but it is not clear whether the same effect can be observed in atomistic datasets.
If true, this enables us to: (1) explain trends in learning curves in ML potentials; (2) quantify redundancy in existing datasets; and (3) evaluate the sampling efficiency of iterative dataset generation methods (Fig. \ref{fig:01-overview}a).
As an initial test for (1), we computed the entropy $\Info$ as a function of dataset size of different molecules in the rMD17 dataset,\citeMain{Chmiela2017} which has been widely used to evaluate the performance of different MLIPs.
The bandwidth was set to a constant value of 0.015 \AA$^{-1}$ to ensure that data points have small overlap, which represents an underestimation of the extrapolation power of MLIPs (see \methods\ for more detail on the choice of bandwidth parameter).
The information entropy of three examplar molecules is shown in Fig. \ref{fig:02-dset}a (Fig. \ref{fig:si:04-rmd17-all-entropies} for all molecules).
In the low data regime, the total dataset entropy increases rapidly with the number of samples.
On the other hand, in the high data regime, the values of $\Info$ saturate because little novelty is obtained from more data points sampled from the same MD trajectories.
As expected, the saturation point depends on the molecule under analysis.
Benzene, a stiff molecule with six redundant environments for both carbon and hydrogen, reaches its maximum entropy in less than 100 samples.
Azobenzene, a molecule with atomic environments exhibiting two- and four-fold degenerate environments, is nearly at its maximum entropy value at 1000 samples.
Although our method does not consider element types, elements may sometimes be distinguished by their environments (i.e., valence rules) and diversity of vibrational motion, allowing the (approximate) quantification of information entropy in a trajectory.
The information on the MD trajectory of aspirin, a much more diverse molecule, is not fully converged even at 10,000 samples.
As this molecule has more rotatable bonds and unique atomic environments than its counterparts, it is expected that its information content is larger, as shown by its higher entropy, and requires more samples to saturate.

We hypothesize that the mismatch between the amount of information in each molecule and the constant number of samples used can partially explain the trends in testing errors across models.
To validate this observation, we compared the information gap --- defined as the information entropy difference between the asymptotic and finite sample size values in Fig. \ref{fig:02-dset}a  --- with the testing errors reported for MACE models trained on these per-molecule dataset splits.\citeMain{Batatia2022}
The correlation between the two quantities is shown in Fig. \ref{fig:02-dset}b, and the information gap curves are shown in Fig. \ref{fig:si:04-rmd17-info-gap}.
The information gap is a strong predictor of the error in forces, with a Pearson correlation coefficient of 0.89.
Even for a constant number of samples (Fig. \ref{fig:si:04-rmd17-const}), the information gap explains major variations in force errors for the models, with the ethanol molecule being the only exception to the trend (see \supptext, Sec. \ref{sec:si:rmd17} and Figs. \ref{fig:si:04-rmd17-energy-hist},\ref{fig:si:04-rmd17-energy-std}).
This suggests that, in a typical MLIP model, the information gap may relate to a minimum theoretical error that can be achieved across a sampled  PES, similar to the lossless compression theorem for information theory.
Conversely, test errors for molecules such as benzene may be equivalent to the training error of the models, as a near-zero information gap suggests that the training set contains complete information about a given configuration space.
As the benchmarks in the literature are performed at a constant number of samples, test errors vary due to differences of information content in each subset, and the information metric can be used to create trade-offs between accuracy and training set sizes.

\subsubsection{Dataset completeness in solid-state systems}

Analogously, this notion of completeness can be useful to post-process existing datasets and quantify redundancy due to sampling and data curation.
Within information theory, entropy is used to guide the development of lossless compression algorithms and encoding methods, which is closely related to our dataset reduction goal without loss of information.
To demonstrate this approach beyond the rMD17 molecular dataset, we computed the entropy of different subsets of the carbon GAP-20 dataset,\citeMain{rowe2020accurate} herein referred to as GAP-20.
The comparison between the subset entropy and the maximum possible entropy for a dataset with the same number of environments is shown in Fig. \ref{fig:02-dset}c (see Fig. \ref{fig:si:04-gap20-entropy}).
This difference between the maximum possible entropy and the subset entropy, shown with grey bars in Fig. \ref{fig:02-dset}c, is complementary to the information gap.
Instead of quantifying how much information is needed to reach a converged dataset, a large difference between $\log n$ and the dataset entropy often indicates oversampling in a dataset.
In the field of MLIPs, test errors are typically used to quantify saturation of a dataset.\citeMain{rowe2020accurate}
However, our information theoretical analysis provides absolute bounds to the entropy and quantifies the completeness of the dataset without training any model.
For example, the difference between the actual information contained in the ``Graphene'' subset of carbon GAP-20 and the absolute limit given by $\log n$ shows that this subset has large redundancy compared to the ``Fullerenes'' subset, where the difference between the maximum and actual entropy is smaller.
The bounds also illustrate how different datasets can exhibit larger diversity.
For example, structures labeled under the ``Liquid'' and ``Amorphous Bulk'' categories are maximally diverse (Fig. \ref{fig:si:04-gap20-entropy}), with environments mostly distinct with the bandwidth used to compute the KDE (0.015 \AA$^{-1}$, see \methods).
This may be a consequence of both the larger accessible phase space by these amorphous and liquid structures and the original farthest point sampling approach used when constructing the dataset.\citeMain{rowe2020accurate}

To illustrate the relationship between information entropy and dataset compression, we computed the entropy curves of different subsets of the GAP-20 dataset.
Then, we trained a NNIP based on the MACE architecture\citeMain{Batatia2022} on randomly sampled fractions of the subsets,  computing test errors as a function of training set size and, thus, entropy.
Fig. \ref{fig:02-dset}d exemplifies this relationship for the labels ``Graphene'' and ``Fullerenes'' of GAP-20, which exhibit large (Graphene) and small (Fullerenes) levels of redundancy (Fig. \ref{fig:02-dset}c).
In the former, datasets as small as 20\% of the original one still exhibit entropies around 4.25 nats, similar to the full one.
Accordingly, their test errors remain constant across all dataset sizes (Fig. \ref{fig:02-dset}e), with a value of 0.96 $\pm$ 1.37 meV/\AA~ for force errors relative to model trained on the full training set.
Despite fluctuations in total entropy caused by the random sampling approach --- which depend on unit cell sizes and ordering of structures in the dataset, and become more sensitive at the low-data regime --- these results show that our model-free analysis of dataset entropy correctly informed the redundancy of the dataset.
On the other hand, the dataset labeled as ``Fullerenes'' is less redundant, and subset entropies monotonically decrease as the training set size goes down.
As expected, the test errors also increase with smaller training set sizes, reproducing known patterns in learning curves of MLIPs (Fig. \ref{fig:02-dset}e).
Although this example considers only a random sample of data points when ``compressing'' a dataset, different algorithms can be used in future work to evaluate optimal subsets with maximum entropy for compression of training sets for MLIPs.\citeMain{li2023exploiting}
The method can also be used to evaluate extrapolation conditions or dataset completeness when fast data generation approaches are targeted.\citeMain{schwalbe2023controlling}

\subsubsection{Information efficiency and diversity in active learning loops}

Finally, to exemplify how information theory can be useful to evaluate active learning (AL) strategies in MLIP-driven atomistic simulations, we analyze dataset metrics of the ANI-Al dataset,\citeMain{Smith2021} which is a dataset for aluminum constructed by starting from random structures and performing over 40 generations of sampling and retraining with NNIP-driven MD simulations.
Figure \ref{fig:02-dset}f shows how the entropy varies as new configurations are sampled by the AL.
In the initial stages of the active learning, the entropy of the dataset quickly increases, then peaks around generation 12, before subsequently decreasing.
To explain this effect, we observe that the increase in  diversity of this dataset comes at the cost of oversampling certain regions of the configuration space.
In fact, fewer than 5\% of the environments sampled after the third round of AL are novel according to our information-theoretical criterion (Fig. \ref{fig:02-dset}g).
This agrees with the known fact that although MD simulations provide a physically meaningful way to sample new configurations, most sampled configurations are correlated and may be already contained in the original training sets.
This may be especially true for large periodic cells where a handful of unknown environments (i.e., $\dH > 0$) may not be easily separated from the numerous known (or similar-to-known) environments ($\dH \leq 0$) that may surround them.
To verify that the total coverage of the configuration space still increases, we propose an additional metric of dataset diversity $D$,

\begin{equation}
    D\left(\Xset\right) =
    \log \left[
    \sum_{i=1}^n e^{\dH(\X_i)}
    \right],
\end{equation}

\noindent that reweights each data point's contribution to the information entropy based on how well-sampled its region of the configuration space is (\supptext, Section \ref{sec:si:diversity}).
Indeed, Figure \ref{fig:02-dset}f shows how the dataset diversity continues to grow even when the entropy decreases.
This approach of measuring dataset diversity is related to the concept of ``efficiency'' in information theory\citeMain{shannon1948mathematical} and may be used to propose new ways to sample atomistic configurations or automatically create datasets for MLIPs in the future.

\subsection{Model-free uncertainty quantification for machine learning potentials}
\label{sec:uq}

When information theory is used to analyze a single dataset, as in the previous section, environments are compared against other environments in the same dataset.
However, it is convenient to test the case when reference datasets $\Xset$ do not contain a tested sample $\Y$, often leading to $\dH (\Y | \Xset) > 0$.
In this scenario, we propose that \takeaway{differential entropies can be used as a model-free uncertainty estimator for a given dataset}.
Whereas uncertainty quantification (UQ) methods for MLIPs usually rely on models\citeMain{hu2022robust,tan2023single} --- i.e., prediction uncertainties are associated to variances in model predictions --- we propose instead that an estimate for UQ in MLIPs can be obtained from the data alone.
This approach is similar to Gaussian process regression methods,\citeMain{Bartok2010,glielmo2018efficient,Vandermause2020} which compute an uncertainty by inverting a covariance matrix computed for training points, or parametric models on a latent space.\citeMain{zhu2023fast}
Differently from other approaches, however, our method performs a fast non-parametric estimate directly on the atomistic data space, thus bypassing the need for a model.
While this approach can be expensive for large datasets, it is easily parallelizable and is guaranteed to provide a robust uncertainty estimate, as it does not rely on the randomness associated with model training or inference.
Rather, it provides a conservative, yet deterministic way to compute uncertainties by calculating the relative information content of a data point using the training set as a reference, while making no assumptions on the extrapolation power of the model that was trained on these data points.
Therefore, our UQ method  can be useful to provide rigorous bounds to epistemic uncertainty regimes, where model-based UQ approaches typically struggle, and to complement model-based UQ when a more clear understanding of model behavior is expected.

\begin{figure}[!h]
    \centering
    \includegraphics[width=\linewidth]{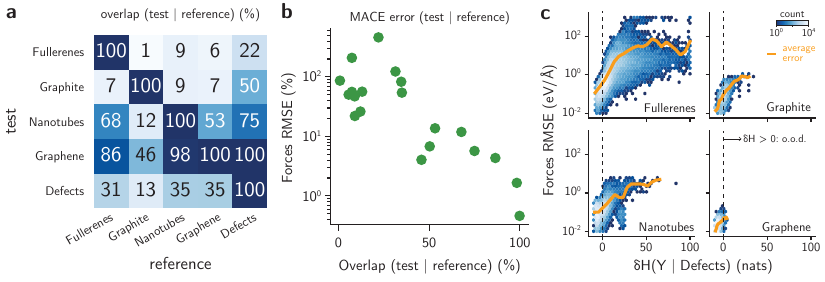}
    \caption{
    \takeaway{Information entropy quantifies  overlaps between datasets and is a model-free UQ method.}
    \textbf{a}, Overlap between test and reference sets for the GAP-20 carbon dataset.
    Only a subset of the data is shown for clarity (see Fig. \ref{fig:si:04-gap20-dH-table} for complete matrix).
    \textbf{b}, Test errors (in \%) of a MACE model trained on one of the subsets of the GAP-20 dataset shown in \textbf{a}, and tested on the other subsets.
    Each point corresponds to a single (train, test) pair.
    Models with higher overlaps between train and test sets exhibit substantially smaller errors.
    \textbf{c}, Correlation between force errors (in eV/\AA) and $\dH$ for models trained on the ``Defects'' subset of GAP-20.
    The average RMSE (orange line) increases with higher $\dH$.
    For ``Fullerenes'', The $\dH$ was truncated to 100 nats for clarity.
    }
    \label{fig:03-uq}
\end{figure}

To exemplify how information theory can be used for UQ in MLIPs, we computed the values of $\dH$ for different subsets of the GAP-20 dataset discussed in the previous section.
Then, we computed the overlap between one subset given another reference set of configurations by quantifying the fraction of points with $\dH (\Y | \Xset) \leq 0$.
Figure \ref{fig:03-uq}a exemplifies these overlaps for the ``Fullerenes'', ``Graphite'', ``Nanotubes'', ``Graphene'', and ``Defects'' subsets of the dataset (see Fig. \ref{fig:si:04-gap20-dH-table} for complete results).
The results show that environments in the ``Graphene'' split are mostly contained in the other subsets, with a minimum overlap of 46\% between ``Graphene'' and ``Graphite''.
Though counter-intuitive from the physics perspective, the ``Graphite'' dataset contains bulk structures which, from the atomic environment perspective, are farther from ``Graphene'' compared to, say, ``Nanotubes'' or ``Fullerenes.''
Interestingly, ``Nanotubes'' contains almost all environments of the ``Graphene'' dataset, but ``Graphene'' contains only 53\% of the environments in ``Nanotubes.''
Similarly, ``Fullerenes'' contains a sizeable portion of ``Nanotubes'', with an overlap of 68\%, but not the other way around.
This analysis also allows us to identify how each subset is constructed without having to label the structures beforehand.
For example, Fig. \ref{fig:si:04-gap20-dH-table} shows that the ``Graphene'' subset is also contained by the  ``Defects'' and ``Surfaces'' datasets, but not fully covered by the (bulk-like) ``Graphite'' dataset.
The subsets labeled as amorphous or liquid do not overlap with any of the others, even though their phase space could have been similar depending on their construction method.
Finally, large subsets such as ``Defects'' and ``SACADA'' contain several parts of the other subsets, largely due to the way they were created.
While the discussion here is based on (human) labels attached to subsets of the GAP-20 dataset, this overlap analysis can be used to compare arbitrary pairs of datasets in general, regardless of available labeling. For example, the similarity of test/validation splits can be analyzed to verify whether small errors are due to good model performance in generalization tasks (small overlaps) or simply because the splits are overly similar to the original training set (high overlaps) without any prior information on the structures.

To verify whether overlap between training and testing sets is useful as a predictor of uncertainty and error metrics, we trained models based on the MACE architecture to each one of the subsets of GAP-20 in Fig. \ref{fig:03-uq}a, then tested the models on the other splits.
Figure \ref{fig:03-uq}b shows the test errors obtained from such training-testing splits.
When models are tested on subsets exhibiting large overlap with their training set, all of them perform well, with normalized errors below 10\%.
On the other hand, errors tend to be much larger when the overlap between train and test sets are small, sometimes surpassing 100\% error (see also Fig. \ref{fig:si:03-uq-parity}).
These results show a power law for distinct train/test sets with clear anti-correlation between the error and overlap.

To further this observation, we evaluated the MACE model trained on the ``Defects'' split of the GAP-20 dataset against the other other four splits, exhibiting increasing overlaps with ``Defects'': ``Fullerenes'' (22\% overlap), ``Graphite'' (50\%), ``Nanotubes'' (75\%), and ``Graphene'' (100\%) (Fig. \ref{fig:03-uq}a).
Force errors were then evaluated for this model and correlated with the values of $\dH$, for each environment, as shown in the distributions of Fig. \ref{fig:03-uq}c.
For environments where $\dH > 0$, the RMSE is often above 0.1 eV/\AA.
On the other hand, when $\dH \leq 0$, errors are typically lower than that.
To demonstrate that higher values of $\dH$ usually lead to higher errors beyond the correlation plots, we computed the average RMSE for each window of $\dH$.
Figure \ref{fig:03-uq}c shows that average errors continue to increase as the values of $\dH$ also increases, showing that points further away from the training set tend to exhibit higher errors.
On the other hand, points slightly outside of the known domain, thus with positive but near zero $\dH$, often show average errors comparable to the ones in the training set.
Interestingly, Fig. \ref{fig:03-uq}c also shows that force errors continue to decrease as $\dH$ becomes more negative.
This correlates with the idea that unbalanced datasets bias the training process and end up minimizing the loss for data points with higher weight (i.e., with more negative $\dH$).
The same observation is valid for the maximum error within each range of $\dH$ (Fig. \ref{fig:si:03-uq-defects-dH}), illustrating how the differential entropy does not exhibit false negatives for the dataset and model under study, i.e., negative entropy values necessarily lead to small errors provided that errors are small everywhere in the training set.
Furthermore, because the uncertainty threshold $\dH > 0$ is guaranteed by the theory to denote extrapolation (\supptext, Section \ref{sec:si:dH}), our UQ metric detects points outside of the training domain without the need for additional calibration or empirically fitted parameters.
Thus, our information theoretical approach provides a robust, model-free alternative to quantifying errors in MLIPs and also can be used beyond NN models.

\subsection{Information theory explains chemical and error trends across the TM23 dataset}

\begin{figure}[!htb]
    \centering
    \includegraphics[width=\linewidth]{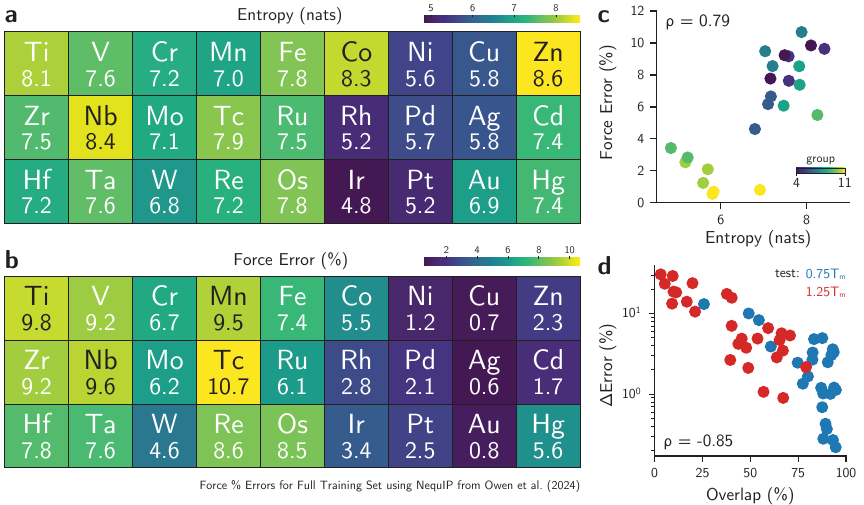}
    \caption{
    \takeaway{Information theoretical quantities correlate with error and chemical trends in the TM23 dataset.}
    \textbf{a}, Information entropy of the full TM23 training set for each element.
    \textbf{b}, Force errors (in \%) for NequIP models trained on the full training set, obtained from Owen \textit{et al.}
    \textbf{c}, These two quantities exhibit strong correlation, as indicated by the Pearson correlation coefficient of $\rho = 0.79$ for transition metals with incomplete d-shell.
    \textbf{d}, The difference between the final forces error (in \%) and the initial forces errors (in \%) (denoted as $\Delta$Error) is explained by the dataset overlap obtained by computing $\dH (0.75 T_m | 0.25 T_m)$ or $\dH (1.25 T_m | 0.25 T_m)$, as demonstrated by the Pearson correlation coefficient of $\rho = -0.85$.
    Red and blue dots indicate error differences for models trained on the ``cold'' subset of TM23 (sampled at 0.25 $T_m$, where $T_m$ is the melting temperature) and tested on the ``warm'' (0.75 $T_m$) and ``melt'' (1.25 $T_m$), respectively.
    }
    \label{fig:05-tm23}
\end{figure}

To demonstrate that the methods above can be combined to analyze other datasets, we used our information entropy formalism to explain trends in the TM23 dataset,\citeMain{owen2024complexity} a newly-proposed benchmark containing structures and properties of elemental transition metals across temperatures.
Owen \textit{et al.} demonstrated that trends in force errors of models fitted to separate elements in this dataset are challenging to explain from elemental properties alone and are better described by many-body interactions due to the electronic structure of the metals.\citeMain{owen2024complexity}
We show that these trends in electronic structure and model errors agree with the theoretical analysis in this work by quantifying the entropy, diversity, information gap, and dataset overlaps for each transition metal in the TM23 dataset.
Figure \ref{fig:05-tm23}a depicts the information entropy for each elemental subset of TM23 using the ``full'' data split, i.e., including information about all three temperatures of the dataset.
The entropy table, arranged like the d-block of the periodic table, indicates with brighter colors the elemental subsets of TM23 with higher information entropy.
The values immediately resemble the original relative force errors for NequIP models trained to the full dataset,\citeMain{owen2024complexity} shown in Fig. \ref{fig:05-tm23}b, where coinage metals exhibit smaller entropy and smaller errors, and early transition metals display higher information entropy and higher force errors (see also Figs. \ref{fig:si:05-tm23-tables}--\ref{fig:si:05-tm23-corr}).
At the same time, Zn, Cd, and Hg are outliers not only in force errors, as pointed out by Owen \textit{et al.},\citeMain{owen2024complexity} but also contain higher information entropy.
As entropy and diversity are related, Fig. \ref{fig:si:05-tm23-diversity}a shows that more diverse datasets often lead to higher relative errors in force prediction, with these quantities having a Pearson correlation coefficient of $\rho = 0.74$.
These results are also correlated with the d-band center of the metals (excluding group 12), showing that differences in the electronic structure are related to the information contents in the dataset (Fig. \ref{fig:si:05-tm23-diversity}b).
As the electronic structure and the accessible phase spaces are strongly connected, our information entropy is measuring the indirect consequence of diversity in configurations that, in essence, are governed by the electronic structure of the materials.
Qualitative visualization of Figs. \ref{fig:05-tm23}a,b may also explain why group VI elements exhibit a lower error compared to group V or VII ones, with the information entropies recovering smaller information contents in Mo compared to Nb or Tc, and similarly for W compared to Ta or Re.
To verify if dataset incompleteness (i.e., low number of samples) also impacted the force errors in the TM23 dataset, we computed the information entropy as a function of the number of environments in the training set, as exemplified in Fig. \ref{fig:si:05-tm23-learning}, and found the information gap to be strongly correlated with the diversity and entropy of the datasets (Fig. \ref{fig:si:05-tm23-corr}).
This finding recovers the expected trends of information-rich datasets being ``harder to learn'' (e.g., Os or Ti), while datasets with less information such as Cu or Ag exhibit smaller errors.
Thus, more diverse phase spaces may require more samples to ensure lower information gaps and (possibly) lower errors.

To study model generalizability, the authors of the TM23 dataset also proposed a test of model transferability across temperatures using the sampled structures.
Specifically, models trained on low-temperature samples (subset ``cold'') are tested against high-temperature samples (subsets ``warm'' and ``melt''), and the resulting errors are compared.
Interestingly, the original benchmark demonstrated that different elemental datasets exhibit varying levels of transferability, with good extrapolation behavior in platinum-group metals and coinage metals, but poor transferability in early transition metals such as Re or Cr.
Using the information-based UQ method described in Section \ref{sec:uq}, we explain these trends in model performance solely from the analysis of dataset overlap using the differential entropy $\dH$.
Using the low-temperature samples as reference dataset, we computed the values of $\dH$ (warm $|$ cold) or $\dH$ (melt $|$ cold) for each element, and computed the overlap by measuring the fraction of the dataset with $\dH \leq 0$.
The comparison between error differences --- i.e., the degradation of the model performance relative to the model trained on the full elemental dataset --- and the dataset overlaps is shown in Fig. \ref{fig:05-tm23}d.
The strong anti-correlation between the error differences and the dataset overlaps, with a correlation coefficient $\rho = -0.85$, further confirms the results discussed above for the model-free UQ analysis.
Interestingly, it also explains why transferability trends shift as we move from the ``warm'' (0.75 $T_m$) to the ``melt'' (1.25 $T_m$) datasets.
For instance, Re has the worst error in the transferability test from ``cold'' to ``warm'', as also shown in Fig. \ref{fig:si:05-tm23-overlaps}, but not in the ``cold'' to ``melt'' test, where models trained to the Zr, Hf, and Cr datasets are the worst-performing ones.
Our analysis shows that the overlaps between the ``melt'' and ``cold'' subsets for the latter three examples indeed drop to levels smaller or equal to Re (i.e., nearly zero overlap, see Table \ref{tab:si:05-tm23-overlaps}), whereas the overlap between their ``warm'' and ``cold'' subsets is substantially larger ($\sim$50\%) than the one for Re.
Our analysis also revealed meaningful information on outliers that remained unexplained in the original TM23 results.
For example, Tc has the worst force error (in \%) when models are trained to the full dataset (Fig. \ref{fig:05-tm23}b), but better transferability across temperatures. Our analysis explains these results
 by the higher overlap in phase spaces between the ``cold'' and ``warm'' (92\%) and ``cold'' and ``melt'' (79\%) for Tc compared to the other metals, suggesting that the process of heating and melting when constructing the dataset for Tc did not vary the phase space as much.
Though much more detailed analysis can be performed across the chemical trends, the  results shown here demonstrate that performance in MLIPs can be predicted even in the absence of models, further strengthening the usefulness of the information-theoretical analysis for several tasks in atomistic ML.

\subsection{Information-based detection of outliers and rare events in atomistic simulations}

Our information theoretical method can further be used to detect outliers (on-the-fly) within a large-scale production MLIP-driven MD simulation and is not restricted to only comparing static datasets.
To demonstrate this concept, we produced an MD trajectory of dynamically strained Ta using a supercell containing approximately 32.5 million atoms and a SNAP potential\citeMain{thompson2015spectral} fitted to an EAM potential (see \methods).
This choice of potential was made so we could have access to the ground-truth energies and forces even at very large scales, thus allowing us to benchmark our outlier detection approach.
In such large models, obtaining uncertainty estimates of energy/force predictions can be challenging even at the postprocessing stage, especially if it requires re-evaluating predictions with several models, such as with an ensemble-based UQ approach.
Furthermore, uncertainty thresholds may not be well-defined for models such as SNAP, where the choice of weights, training sets, and hyperparameters can lead to substantial variations of model performance.\citeMain{MontesdeOcaZapiain2022}
Finally, ML-driven simulations of periodic systems may fail in completely different ways compared to simpler molecular systems,\citeMain{Fu2022,vita2023losslands} where bond lengths and angles are sometimes sufficient to detect extrapolation behavior.

\begin{figure}[!htb]
    \centering
    \includegraphics[width=0.9\linewidth]{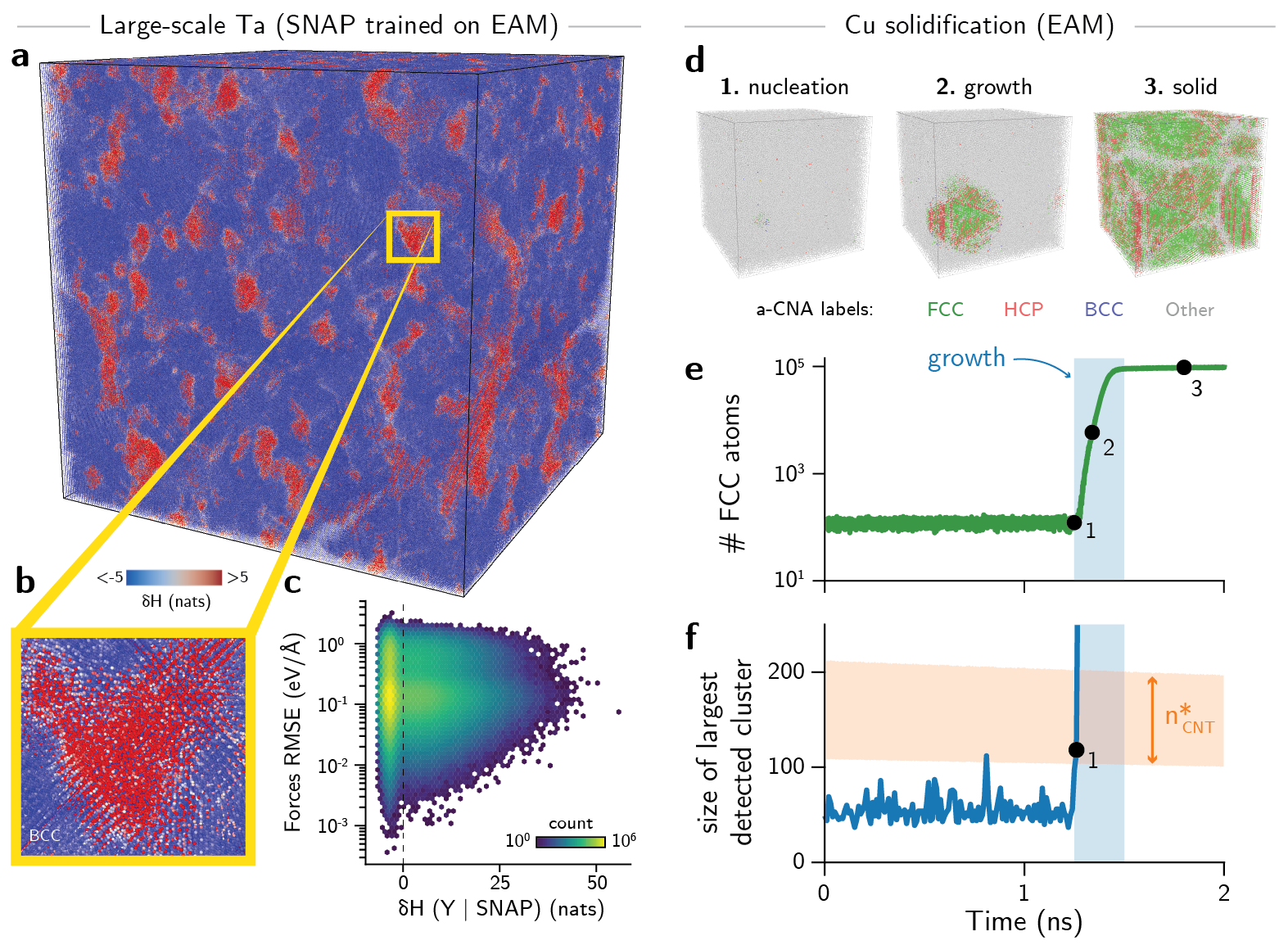}
    \caption{
    \takeaway{Differential entropies detect outliers and rare-events in large atomistic simulations.}
    \textbf{a} Visualization of a 32.5M atom snapshot of BCC Ta simulated with SNAP.
    Colors represent the values of the estimated $\dH$, with blue atoms indicating environments reasonably within the training set ($\dH < 0$) and red atoms indicating environments outside of the training set ($\dH > 0$).
    Values of $\dH$ were truncated to the range $[-5, 5]$ to facilitate the visualization of divergent colors.
    \textbf{b}, Example of high-uncertainty region encountered during the simulation. The formation of a disordered, non-BCC phase (red) in the simulation leads to unphysical behavior in the trajectory.
    \textbf{c}, The unphysical behavior cannot be identified only by errors in forces. Even outside of its known domain, the SNAP model exhibits errors within the range of systems within the training set.
    The number of environments in each region is shown by the color scale, with brighter colors indicating exponentially denser regions of the error-$\dH$ space.
    }
    \label{fig:04-outlier}
\end{figure}

Using the values of $\dH$ computed for each environment (i.e., atom center) in the 32.5M atom system against the original training set, we analyzed a snapshot of the Ta MD trajectory with our information theoretical method to identify possible anomalies due to extrapolation during the simulation.
Figure \ref{fig:04-outlier}a shows that about 13\% of the environments at this time step exhibit $\dH > 0$ (red atoms), some of which are as large as $\dH = 55.8$ nats (see Fig. \ref{fig:si:05-Ta-density}), showing that a substantial number of environments are outside of the training domain of the potential.
This particular simulation was initialized with a monocrystalline BCC crystal of Ta seeded with dislocations, following the approach from Zepeda-Ruiz \textit{et al}.\citeMain{zepeda2017probing}
After propagating the trajectory with MD simulation, much of the cell retained BCC character, which is well represented in the training set (and thus appears as the blue colored atoms), but the simulation also proceeded to form amorphous-like phases (Fig. \ref{fig:04-outlier}a,b) that are unexpected in such trajectories and are detected as the out-of-domain outliers (red).
Although the model prevents obvious nonphysical configurations, such as overlapping atoms, distinguishing between such model failures and new physical phenomena (such as finding reasons for strain hardening) in these simulations is challenging without our information theoretical approach.
To illustrate the power of this outlier detection and the difficulty of otherwise \textit{a priori} identifying a possible model failure, we computed the ground truth forces for the atomistic system from the interatomic potential that originated the labels for the SNAP training set, and analyzed the errors of the predictions as a function of $\dH$ for each configuration (Fig. \ref{fig:04-outlier}c).
The results show that the SNAP model under investigation does not exhibit high extrapolation errors, as measured by the RMSE of forces, which are within the same range of errors for environments having $\dH > 0$ and $\dH < 0$.
Instead, the formation of the amorphous phase is likely due to lower predicted energies compared to the true energies for these out-of-domain configurations.
Therefore, even access to the ground truth forces does not allow the classification of a trajectory as failed within these constraints, and instead would rely on human inspection of the trajectory, which can be subjective and is prone to observer error. We note that the unexpected behavior demonstrated in this case appeared late in the trajectory and is extremely challenging to anticipate without outlier detection.
On the other hand, our differential entropy detects these outliers without the need for a calibrated threshold, providing a conservative estimate for understanding extrapolation in a model-free approach, and providing ``early warning'' that can be used for model augmentation, re-evaluation, or retraining.

Beyond ML robustness and UQ, finding outliers in atomistic simulations is also of importance for simulations involving phase transitions and other out-of-equilibrium events, and our approach can be used for trajectory and physical analysis.
Because our information-theoretical method is model-independent, we illustrate in particular how the approach can be used to detect rare events in MD trajectories.
As a model system, we analyzed the nucleation of copper simulated using the potential from Mishin \textit{et al.}\citeMain{mishin2001structural} in a supercell containing nearly 300,000 atoms with undercooling of approximately 420 K at 1 bar pressure.
The main stages observed during the trajectory included: (1) nucleation of a crystal from the melt; (2) crystal growth regime; and (3) solidified system with residual liquid and grain boundaries (Fig. \ref{fig:04-outlier}d).
The nucleation event was obtained by gradually decreasing the temperature of the molten system over 2 ns (Fig. \ref{fig:si:04-solid-curves}).
Classification of the atomic environments using the traditional common neighbor analysis (CNA)\citeMain{faken1994systematic} reveals the appearance of a dominant FCC phase in the second half of the simulation, with rapid growth for about 100 ps, and a plateau in later stages (Fig. \ref{fig:04-outlier}e).

This process of nucleation and growth is typically explained by approaches such as the classical nucleation theory (CNT), which uses near-equilibrium assumptions to model an out-of-equilibrium event.
The CNT proposes that there is a statistical distribution of solid clusters in the liquid phase, and the nucleation event happens when an outlier cluster of critical size is formed in a probabilistic event.
However, observing or detecting these outlying critical nuclei directly from atomistic simulations can be challenging.
Discrete classification of the solid phases, e.g., using variations of the CNA, typically prevents identification of the structure of the liquid phase, as very few clusters are found and classified correctly as solid-like when they are found in the melt.
To verify if our information theoretical model can detect outliers as in nucleation theory, we computed the values of $\dH$ for each frame in the simulation using an MD trajectory of pure FCC copper as reference.
To ensure a conservative estimate of what defines a ``solid-like'' cluster, the reference dataset was taken from a MD simulation at a low temperature of 400 K and pressure of 1 bar in the NPT ensemble.
Then, frames of the original solidification trajectory were compared against snapshots of this low-$T$ reference MD.
Instead of considering environment labels assigned by an algorithm, we detected nuclei in the melt by finding environments with high overlap with the phase space of the pure solid.
In practice, this means that subcritical nuclei can be identified by taking environments $\X_\mathrm{melt}$ such that $\dH(\X_\mathrm{melt} | \{\X_\mathrm{solid}\}) < 0$.
Using this method, we obtained the number and size of clusters by using a graph theoretical analysis, where nodes are environments with $\dH \leq 0$, edges connect nodes at most 3 \AA\ apart, and clusters are the sets of connected components (\methods).
Then, we analyzed the largest cluster size among all extracted subgraphs (Fig. \ref{fig:04-outlier}f).
With this analysis, we observed the largest cluster size is typically below 100 atoms until the nucleation event, when it reaches the value of 114 atoms.
In contrast, the CNA method recovers a maximum of 170 FCC-like environments within the entire simulation box and across all pre-nucleation frames.
To compare this with predictions from the CNT, we calculated the critical nucleus size given the average, time-dependent undercooling in the simulation.
The melting enthalpy and temperature obtained from the simulation's EAM potential were used to perform such estimate,\citeMain{mendelev2010molecular} and the solid-liquid interfacial energy was taken from experiments, which range between 0.177 and 0.221 J/m$^2$ for Cu.\citeMain{turnbull2004Formation,vinet2002Correlations,kaptay2020Coherent}
This range of predictions is shown in Fig. \ref{fig:04-outlier}f in orange color.
The results demonstrate that nucleation happens when the largest cluster identified by our information theoretical method falls roughly within the range of experimental critical nucleus sizes.
Prior to the nucleation event, only a single other frame intersects the region of maximum cluster size.
Visualization of the cluster indicates that the graph at that frame is better approximated as two nuclei rather than a single, compact critical nucleus (Fig. \ref{fig:si:04-nuclei}, see also Sec. \ref{sec:si:cnt-clusters}).
This demonstrates that our method can be used to reliably and sensitively detect rare events in atomistic simulations.

\section{Discussion}

Our results show how an information theoretical approach can be used in a range of problems in atomistic modeling.
By computing distributions of atom-centered representations from simulations, our information-based analysis of datasets and uncertainty explains multiple results within learned interatomic potentials and atomistic simulations in general, with a fast model-free approach.
In particular, we showed how information and diversity content in a dataset can be predicted directly from a training set, explaining error trends in MLIPs, rationalizing dataset compression, predicting extrapolation errors, and detecting failed simulation trajectories.
Moreover, because information entropy provides a quantitative estimation of ``surprise'' of a random variable, we proposed its use as a robust UQ metric for ML-driven atomistic simulations and showed it can be used for outlier detection even for large atomistic systems.
As this strategy does not depend on models, it can be adapted to any MLIP (or any fitted potential) to provide objective uncertainty metrics and can be used as a general UQ method for atomistic simulations. Similarly, the method was demonstrated to successfully detect of rare events.

Interestingly, beyond its use in MLIPs and atomistic datasets, we observed that the information entropy of atom-centered descriptor distributions qualitatively recovers the configuration component of thermodynamic entropy differences  (i.e., entropy differences due to vibration, but not momenta or composition) in several cases.
For example, we obtained surprisingly accurate predictions of entropy differences for several elemental systems by adjusting the bandwidth parameter according to a physics-based rule, allowing us to compute phase diagrams at high pressures and temperatures (see \supptext\ \ref{sec:si:thermo-sec}, Figs. \ref{fig:si:01-bandwidth} and \ref{fig:si:02-thermo}).
Whereas the parallels between information and thermodynamic entropies are well-known, there is no guarantee that descriptor distributions should represent the actual thermodynamic entropies beyond simple configurational components.
Nevertheless, the coincidence between information and the configurational component of thermodynamic entropies will be investigated in more detail in upcoming works.

In the future, several strategies can further generalize this method.
For example, the approach does not explicitly account for element types due to the choice of representation.
For simple molecular systems, bonding patterns (e.g., valence rules) sometimes map distributions of atomic environments to different parts of the information entropy space due to the construction of the $k$-nearest neighbors descriptor based on interatomic distances.
However, for inorganic crystals, this approximation may not be valid, and may have to be incorporated into the approach to account for true configuration entropies, as in alloys.
Furthermore, as the feature space of several atom-centered representations may not be injective,\citeMain{pozdnyakov2020incompleteness} the information entropy depends on the choice of descriptor.
Nevertheless, the formalism is general and can be immediately adapted to other representations by simply changing the metric space and kernel function to other appropriate choices.
The explicit dependence of the information entropy for atomic environments according to the kernel function and descriptors will be investigated in future work.
Finally, whereas the current computational implementation is sufficient for the analysis of tens of millions of environments, improvements in parallelization and hardware utilization can allow the approach to scale towards a real-time UQ for MD.
Our analysis of the parallelized and approximated results for the large-scale Ta system shows that the information entropy can be computed for large systems even with few resources (see \supptext, Sec. \ref{sec:si:approx-dH}).
While computation of kernel matrices already have been  implemented in the code, faster computation of descriptors, multi-node parallelization, or better approximate nearest neighbors computation can be implemented in future versions of the code, allowing greater scaling in computing kernel density estimates and their resulting information contents.

\section{Conclusions}

In this work, we proposed a strategy for quantifying information in atomistic simulations and ML based on information theory.
By performing a kernel density estimation over distributions of atom-centered features, we obtained values of information entropy that:
(1) rationalize trends in testing errors for machine learning potentials across multiple datasets, relating model performance to information contents;
(2) proposes an approach to quantify compressibility and sampling efficiency for atomistic datasets based on information theory;
(3) provides a model-free uncertainty quantification approach for atomistic ML; and
(4) allows for outlier detection in large production simulations from MLIPs or general atomistic datasets.
These contributions are demonstrated with numerous examples, from known benchmarks from the MLIP literature, a solidification trajectory, and a simulation of a system containing about 32.5M atoms.
As increasingly accurate and scalable ML models are proposed for atomistic simulations, this work proposes a rigorous way to optimize their training process, automate evaluation of information contents in datasets, and assess the performance of the models against well-defined theoretical bounds.
Additional developments in atomistic information theory can continue to translate developments in machine learning and other statistical methods into faster and more accurate materials modeling.

\section*{Methods}
\customlabel{sec:methods}{Methods}

\subsection*{Information entropy and QUESTS method}

\noindent\textbf{Representation:} the representation of atomic environments was computed as described in Section \ref{sec:repr} of the main text and Section \ref{sec:si:repr} of the \supptext.
Throughout this work, a number of $k = 32$ neighbors was used to represent the atomic environment, with a cutoff of 5 \AA.
This range of neighbors and cutoffs approximates the hyperparameters typically used in MLIPs, where cutoffs of 5--7 \AA\ are employed.
Whereas changing the values of $k$ and cutoff influences the metric space by making environments increasingly similar (at lower $k$ or radii) or dissimilar (at higher $k$ or radii), we observed that the trends in entropy remained consistent throughout the examples in this manuscript (see also mathematical derivation in the \supptext, Sec. \ref{sec:si:repr}).
To accelerate the calculation of the representation, the code that computes the descriptors was optimized using Numba\citeMain{lam2015numba} (v 0.57.1) and its just-in-time compiler.
For periodic systems, the feature vectors were created by adapting the stencil method for computing neighbor lists and parallelizing the creation of features across bins.

\noindent\textbf{Information entropy:} the information entropy of descriptor distributions was computed as described in Section \ref{sec:repr} of the main text and Section \ref{sec:si:entropy} of the \supptext.
Throughout this work, the natural logarithm was used for the entropy computation, which scales the information to natural units (nats).
The bandwidth was selected by computing the distance between two FCC structures with lattice parameters 3.58 and 3.54 \AA\ (1\% compressive strain), leading to a bandwidth of 0.015 \AA$^{-1}$ (see also Sec. \ref{sec:si:bw} of the \supptext).
For the computation of the differential entropy $\dH$, the bandwidth and units adopted are the same as the information entropy.

\noindent\textbf{Entropy asymptotes:} the asymptotic behavior of entropies in the learning curves of Fig. \ref{fig:02-dset}a and \ref{fig:si:04-rmd17-all-entropies} was obtained by fitting a function of the form

\begin{equation}
    f(N) = a - b \exp \left[
    -c (\log N)^2
    \right],
\end{equation}

\noindent with $a, b, c$ non-negative parameters obtained from the entropy curve as a function of training set size $N$.
The first and last three points were discarded during the fitting process.
The fit was performed using a non-linear least squares method implemented in SciPy\citeMain{2020SciPy} (v. 1.11.1).
This functional form was found to closely approximate the curves shown in Figs. \ref{fig:02-dset}a, \ref{fig:05-tm23}e, \ref{fig:si:04-rmd17-all-entropies},  and \ref{fig:si:05-tm23-learning}.

\noindent\textbf{Entropy of the TM23 dataset:} The entropy was computed for a rescaled dataset to avoid issues with samples at different temperatures and densities, thus facilitating comparisons across metals of different densities using a constant bandwidth of 0.015 \AA$^{-1}$.
The original distributions of atomic volumes (i.e., the reciprocal of densities) for the TM23 dataset is shown in Fig. \ref{fig:si:05-tm23-tables}a.
The final rescaled volume was adopted as the median volume of all elemental datasets, and equal to 15.8 \AA$^3$/atom.

\subsection*{Molecular dynamics simulations}

\noindent All MD simulations were performed using the Large-scale Atomic/Molecular Massively Parallel Simulator (LAMMPS) software\citeMain{Thompson2022LAMMPS} (v. 2/Aug./2023).
All simulations were performed using a 1 fs time step, except when stated otherwise.

\noindent\textbf{Cu solidification:} the solidification trajectory of copper was simulated using the EAM potential from Mishin \textit{et al.}\citeMain{mishin2001structural}
A $42 \times 42 \times 42$ supercell of FCC copper (296,352 atoms) was simulated above the melting point to produce the structure of the liquid, then cooled to 924 K.
Starting at the temperature of 924 K, the system was cooled to 914 K over the course of a 2 ns-long simulation in the NPT ensemble with the Nos\'e-Hoover thermostat and barostat\citeMain{nose1984unified,hoover1985canonical} implemented in LAMMPS.
Damping parameters for the temperature and pressure were set to 0.1 and 3.0 ps, respectively, a 2 fs time step was used for the integrator, and constant pressure of 1 bar.
Over the trajectory, the number of FCC atoms was computed using the common neighbor analysis (CNA) implemented in LAMMPS.\citeMain{faken1994systematic}

\noindent\textbf{Large-scale Ta simulation:} The atomistic configuration with ``amorphous-like'' substructures (Fig. \ref{fig:04-outlier}) used in benchmarking performance of our information-based detection of structural anomalies resulted from a large-scale MD simulation of crystal plasticity in body-centered-cubic metal Ta.
The simulation was performed using a SNAP potential fitted to the dataset of the original SNAP potential.\citeMain{thompson2015spectral}
However, rather than using the DFT ground-truth reference values of energies, forces and stress in the original fitting dataset, all the same quantities were re-computed using an inexpensive interatomic potential of the embedded-atom-method (EAM) type.
Given that both SNAP and EAM simulations can be performed at scales large enough to perform simulations of metal plasticity of the kind described in Zepeda-Ruiz \textit{et al.},\citeMain{zepeda2017probing} the intention was to observe if a SNAP potential fitted to such a proxy training dataset could reproduce plastic strength predicted by the proxy potential itself.
The SNAP simulation considerably diverged from the proxy EAM simulation both in predicted plasticity behavior and in producing the ``amorphous-like'' regions that never appeared in the proxy EAM simulation.

\subsection*{Machine learning potential}

\noindent\textbf{MACE architecture:}
the ML force fields for GAP-20 in this work were trained using the MACE architecture.\citeMain{Batatia2022}
We used the MACE codebase available at \url{https://github.com/ACEsuit/mace} (v. 0.2.0).
Two equivariant layers with $L = 3$ and hidden irreps equal to \texttt{64x0e + 64x1o + 64x2e} were used as main blocks of the neural network model.
A body-order correlation of $\nu = 2$ was used for the message-passing scheme, and the spherical harmonic expansion was limited to $\ell_\mathrm{max} = 3$.
Atomic energy references were derived using a least-squares regression from the training data.
The number of radial basis functions was set to 8, with a cutoff of 5.0 \AA.

\noindent\textbf{MACE training:}
the MACE model in this work was trained with the AMSGrad variant of the Adam optimizer,\citeMain{kingma2015Adam,reddi2018convergence} starting with a learning rate of 0.02.
The default optimizer parameters of $\beta_1 = 0.99$, $\beta_2 = 0.999$, and $\varepsilon = 10^{-8}$ were used.
The exponential moving average scheme was used with weight 0.99.
In the beginning of the training, the energy loss coefficient was set to 1.0 and the force loss coefficient was set to 1000.0.
The learning rate was lowered by a factor of 0.8 at loss plateaus (patience = 50 epochs).
After epoch 500, the training follows the stochastic weight averaging (SWA) strategy implemented in the MACE code.
From there on, the energy loss coefficient was set to 1.0 and the force loss coefficient was set to 100.0.
The model was trained for 1000 epochs.
A batch size of 10 was used for all models, except in the Defects subset of GAP-20, for which the batch size was adopted as 5.
Each dataset was split randomly at ratios 70:10:20 for train/validation/test.
The best-performing model was selected as the one with the lowest error on the validation set.

\textbf{Error normalization:} in Fig. \ref{fig:03-uq}b, the forces RMSE (in eV/\AA) was normalized by the average force in the test set to account for different distributions of forces in different splits of GAP-20.
This strategy is similar to the original adopted in Fig. \ref{fig:05-tm23}, whose values of normalized error (in \%) were obtained from the original work.\citeMain{owen2024complexity}

\subsection*{Uncertainty quantification}

\noindent\textbf{Novelty of an environment:}
a sample $\Y$ is considered novel with respect to a reference set $\Xset$ if $\dH (\Y | \Xset ) > 0$.
Therefore, the novelty of a test dataset $\Yset$ with respect to $\Xset$ is computed as the fraction of environments $\Y_i \in \Yset$ such that $\dH (\Y_i | \Xset ) > 0$.
On the other hand, the overlap between a test dataset $\Yset$ with respect to $\Xset$ is the fraction of environments $\Y_i \in \Yset$ such that $\dH (\Y_i | \Xset ) \leq 0$.
Larger positive values of $\dH$ imply that the test point $\Y_i$ is further away from the training set $\Xset$.

\noindent\textbf{Novelty in active learning:} specifically in Fig. \ref{fig:02-dset}g, the novelty of sampled configurations at generation $n > 1$ is obtained by computing the differential entropy $\dH$ with respect to the complete dataset at generation $n - 1$.

\noindent\textbf{Correlations between error and $\dH$:}
Force errors in Fig. \ref{fig:03-uq}d were computed by taking the norm between the predicted and true force for each atom, thus assigning a single error per environment.
To average the errors for each $\dH$, as shown in Fig. \ref{fig:03-uq}e, we binned the values of $\dH$ in 20 bins of uniform length $\ell$.
Then, for each bin, we averaged the errors for all points within $0.75 \ell$ of the center of the bin.
This creates a running average effect for the errors, reducing the effect of discontinuities with small displacements of bin centers.
At the same time, the bin length $\ell$ is determined by the range of the values of $\dH$.

\subsection*{Classical nucleation theory analysis}

\noindent\textbf{Critical cluster size:} following known results from the classical nucleation theory (CNT), the critical cluster size $r^*$ of a monocomponent, spherical cluster in a melt is given by

\begin{equation*}
    r^* = \frac{2 \gamma_\mathrm{SL} T_m}{\Delta H_m \Delta T},
\end{equation*}

\noindent where $\gamma_\mathrm{SL}$ is the interfacial energy between the solid and liquid, $T_m$ is the melting temperature, $\Delta H_m$ is the latent heat of melting, and $\Delta T$ is the undercooling.
For the solidification of copper, experimental values of $\gamma_\mathrm{SL}$ range between 0.177 and 0.221 J/m$^2$.\citeMain{turnbull2004Formation,vinet2002Correlations,kaptay2020Coherent}
Whereas the experimental melting temperature at 1 bar is  1357.77 K, with latent heat equal to 13.26 kJ/mol, we used the values determined for the potential, with $T_m = 1323$ K and $\Delta H_m = 11.99$ kJ/mol.\citeMain{mendelev2010molecular}
The ranges of critical cluster sizes in Fig. \ref{fig:04-outlier}f were obtained by assuming a spherical cluster and an atomic volume of 12.893 \AA$^3$/atom obtained from the simulations.

\noindent\textbf{Graph-theoretical determination of clusters:}
As classification methods such as (a-)CNA cannot detect solid-like clusters in the melt, we assumed that clusters can be identified by the overlap in phase space between the melt and a pure solid phase.
To create such a reference phase space, we first sampled a trajectory of an FCC Cu solid at 1 bar and 400 K at the NPT ensemble using the potential from Mishin \textit{et al.}\citeMain{mishin2001structural} and the Nos\'e-Hoover thermostat and barostat\citeMain{nose1984unified,hoover1985canonical} implemented in LAMMPS, with damping parameters equivalent to 0.5 and 3.0 ps for the temperature and pressure, respectively.
We simulated a $20 \times 20 \times 20$ supercell containing 32,000 Cu atoms.
Initial velocities are sampled from a Gaussian distribution scaled to produce the desired temperature, and with zero net linear and angular momentum.
The simulation was equilibrated for 40 ps, after which five snapshots separated by 5 ps were saved to create the reference dataset, which contained 160,000 environments.

Using the reference environments, we computed the differential entropy $\dH$ of each frame of the solidification trajectory prior to growth.
Then, we used a graph theoretical approach to determine the cluster sizes.
Specifically, we considered that environments with $\dH \leq 0$ with respect to the solid are nodes in a graph, and edges connect environments at most 3.0 \AA\ apart.
Then, clusters are defined as 2-connected subgraphs of the larger graph.
The cluster sizes are given by the number of nodes in each subgraph, and the maximum cluster in each frame of the trajectory is estimated by the largest subgraph.

\section*{Data Availability}

The datasets used for training/testing ML potentials were obtained from the original sources at:

\begin{itemize}
    \item rMD17: \url{https://figshare.com/articles/dataset/Revised_MD17_dataset_rMD17_/12672038}
    \item GAP-20: \url{https://doi.org/10.17863/CAM.54529}
    \item ANI-Al: \url{https://github.com/atomistic-ml/ani-al}
    \item TM23: \url{https://doi.org/10.24435/materialscloud:6c-b3}
\end{itemize}

\section*{Code Availability}

The code for QUESTS is available on GitHub at the link \url{https://github.com/dskoda/quests}.

\section*{Acknowledgements}
This work was performed under the auspices of the U.S. Department of Energy by Lawrence Livermore National Laboratory (LLNL) under Contract DE-AC52-07NA27344.
The authors acknowledge funding from the Laboratory Directed Research and Development (LDRD) Program at LLNL under project tracking codes 22-ERD-055 and 23-SI-006.
The authors are grateful to the IAP-UQ group at LLNL for useful discussions, T. Hsu for providing the data for the denoised copper trajectories, V. Bulatov for the data on the tantalum simulation, and L. Williams for pointing us to the training set for SNAP.
D. S.-K. additionally acknowledges support from the UCLA Samueli School of Engineering.
Manuscript released as \texttt{LLNL-JRNL-862887-DRAFT}.

\section*{Conflicts of Interest}

The authors have no conflicts to disclose.

\section*{Author Contributions}

\noindent\textbf{Daniel Schwalbe-Koda:} Conceptualization; Data Curation; Formal Analysis; Investigation; Methodology; Project Administration; Software; Validation; Visualization; Writing - Original Draft; Writing - Review \& Editing; Funding Acquisition; Supervision.
\textbf{Sebastien Hamel:} Data Curation; Investigation; Software; Writing - Review \& Editing.
\textbf{Babak Sadigh:} Data Curation; Investigation; Writing - Review \& Editing.
\textbf{Fei Zhou:} Validation; Data Curation; Writing - Review \& Editing; Supervision.
\textbf{Vincenzo Lordi:} Conceptualization; Data Curation; Writing - Review \& Editing; Funding Acquisition; Project Administration; Supervision.


\clearpage
\beginsupplement

\clearpage

\appendix
\customlabel{sec:sinfo}{Supplementary Information}

\bgroup\setlength{\parindent}{0pt}
\begin{flushleft}
  \textbf{\LARGE Supplementary Information for: Model-free quantification of completeness, uncertainties, and outliers in atomistic machine learning using information theory}\vspace*{2.5em}

  \textbf{Daniel Schwalbe-Koda~$^{1, 2}$\footnote{E-mail: dskoda@ucla.edu},  Sebastien Hamel~$^{1}$, Babak Sadigh~$^{1}$, Fei Zhou~$^{1}$, Vincenzo Lordi~$^{1}$\footnote{E-mail: lordi2@llnl.gov}}\\
    \vspace{1em} 
    \normalfont{\small $^{1}$Lawrence Livermore National Laboratory, Livermore, CA 94550, United States}\\
    \normalfont{\small $^{2}$University of California, Los Angeles, Los Angeles, CA 90095, United States}
\end{flushleft}\egroup

\pagenumbering{arabic}
\setcounter{page}{1}

\section{Supplementary Text}
\customlabel{sec:stext}{Supplementary Text}

\subsection{Derivation of the descriptor}
\label{sec:si:repr}

Consider a representation $f: \mathcal{S} \ra \Xspace$ that maps atomic environments $S$ into features $\X \in \Xspace$, with $\Xspace \subset \Real^N$, and denote $f(S_i) = \X_i$.
The effectiveness of the function $f$ is often computed according to the following properties:\citeSupp{Widdowson2022ResolvingData2}

\begin{enumerate}
    \item \textbf{Invariance}: the representation encodes all symmetries of the system, i.e. given a symmetry operation $T: \mathcal{S} \ra \mathcal{S}$, $f(S) = f(T(S))$.
    \item \textbf{Completeness}: if two descriptors are equal, $f(S_i) = f(S_j)$, then the originating structures are equal up to a symmetry operation, $S_i = T(S_j)$.
    \item \textbf{Metric}: the function $f$ induces a metric $d$ in the descriptor space $\Xspace$.
    \item \textbf{Continuity}: arbitrarily small displacements of atoms in $\mathcal{S}$ ideally lead to arbitrarily small distances between features in $\Xspace$.
    \item \textbf{Speed}: the representation should be fast to compute.
    \item \textbf{Invertibility}: given $\X_i = f(S_i)$, it is possible to reconstruct $S_i$ up to a symmetry operation.
\end{enumerate}

The field has many representations, several of which exhibit different properties.
Here, we derive a new representation which is expected to satisfy several of the properties above.
The representation is inspired in simple distances distributions, which have been proven to satisfy these properties \citeSupp{Widdowson2022ResolvingData2} and have been used for other materials systems\citeSupp{schwalbekoda2023inorganic2}.

\subsubsection{Radial terms}

As a first order approximation, one can obtain an invertible mapping between the structures by taking the pairwise distances between atoms, then reconstructing them using the information from all atoms at once \citeSupp{Widdowson2022ResolvingData2}.
In particular, to make a fixed-length representation, one can take the distances towards the $k$-nearest neighbors of each atom as a representation,

\begin{equation}
    r_{ij} = \norm{\mathbf{r}_i - \mathbf{r}_j},
\end{equation}

\noindent where

\begin{equation}
    r_{i1} \leq r_{i2} \leq \ldots \leq r_{ik}.
\end{equation}

\begin{figure}[!h]
    \centering
    \includegraphics[width=0.5\linewidth]{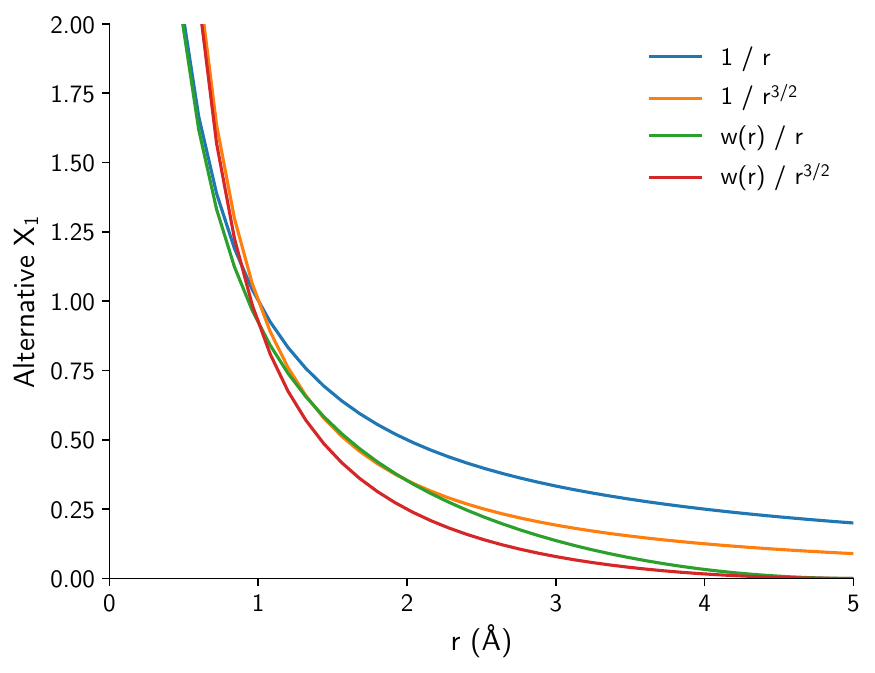}
    \caption{
    Dependence of a proposed $X_1$ functional form according to interatomic distances.
    A cutoff of 5 \AA\ is used for the weight function $w(r)$.
    }
    \label{fig:si:01-x-scaling}
\end{figure}

\noindent As distances between atoms infinitely far apart should be negligible according to a metric that relates to machine learning potentials, we take the representation as being the inverse of distances,

\begin{equation}
    X_{ij}^{(1)} = \frac{w(r_{ij})}{r_{ij}},
\end{equation}

where $X_{i1} \geq \ldots \geq X_{ik}$ and $w$ is a cutoff function given by

\begin{align}\label{eq:si:weight}
    w(r) =
    \begin{cases}
        \left[1 - \left(\frac{r}{r_c}\right)^2\right]^2~ &, r \leq r_c \\
        0~ &, r > r_c
    \end{cases},
\end{align}

\noindent where $r_c$ is a cutoff distance.
The weight function was chosen to satisfy two criteria: (1) fast convergence of the descriptor; and (2) scaling of each component of $\X_1$ approaching $r^{-3/2}$ to  resembles the relationship between entropy and distances in an ideal gas.
Figure \ref{fig:si:01-x-scaling} shows how the combination of the weight function from Eq. \eqref{eq:si:weight} and the inverse distance $1 / r_{ij}$ approximates a dependence of $r_{ij}^{-3/2}$.

In principle, given a large number of neighbors, the unit cell parameters, and $r_c$, an input structure $S$ may be reconstructed from $f(S) = \set{\X^{(1)}}$ up to an isometry. \citeSupp{Widdowson2022ResolvingData2}

\subsubsection{Cross terms}

As the structure can only be reconstructed from the set of representations of all neighbors, increasing the amount of information in each local environment is desirable.
This would also allow us to distinguish between environments containing the same set of nearest-neighbor distances, but different angles.
One way to do so is to incorporate distances between atoms in a neighborhood of $i$,

\begin{equation}
    X_{ijl}^{(2)} = \frac{\sqrt{w(r_{ij}) w(r_{il})}}{r_{jl}},
\end{equation}

\noindent which is performed for each neighbor $l$ of atom $j$ in the neighborhood of $i$.
The weights $w(r_{ij})$ and $w(r_{il})$ ensure that cross distances $r_{jl}$ are less important far away from the center of the neighborhood. The square root ensures that $\mathbf{X}_{i}^{(2)}$ has the same scaling and units as $\mathbf{X}_{i}^{(1)}$.
The final representation on a per-neighbor basis is

\begin{equation}
    \mathbf{X}_{ij}^{(2)} = \left(
        X_{ij1}^{(2)}, \ldots, X_{ijk}^{(2)}
    \right),
\end{equation}

\noindent with the constraint $X_{ij1}^{(2)} \geq \ldots \geq X_{ij(k-1)}^{(2)}$. Finally, the second-order representation term for each atom is given by

\begin{equation}\label{eq:si:repr:main}
    \mathbf{X}_{i}^{(2)} = \frac{1}{k} \sum_j \mathbf{X}_{ij}^{(2)}.
\end{equation}

Where the radial distances $\mathbf{X}_{i}^{(1)}$ already suffice for reconstruction when all $i$'s are considered (along with unit cell parameters),\citeSupp{Widdowson2022ResolvingData2} the pairwise cross distances $\mathbf{X}_{i}^{(2)}$ may help reconstructing environments only from the vector $\X_i = \left(\mathbf{X}_{i}^{(1)}, \mathbf{X}_{i}^{(2)} \right)$, even though reconstruction may not be guaranteed for this descriptor.
If instead of an average in Eq. \eqref{eq:si:repr:main} we concatenated all vectors $\mathbf{X}_{ij}^{(2)}$, then reconstruction could be guaranteed within the sphere limited by $r_c$.
Continuity of this descriptor is only possible when this cutoff $r_c$ is smaller than the distance of the central atom to the $k-$th nearest neighbor, as switching between neighbors would create a discontinuity for the aggregated contributions.
Figure \ref{fig:si:01-descriptors} illustrates the pairwise distances relevant for the construction of this descriptor.

\subsection{Dataset entropy}
\label{sec:si:entropy}

According to information theory, the entropy $\Info$ of a distribution $p(x)$ is defined as

\begin{equation}\label{eq:si:entropy-info}
    \Info = - \int p(x) \log p(x) dx,
\end{equation}

\noindent where $p(x)$ is the distribution of data points, $\log$ is the natural logarithm, and the value of entropy is integrated over the entire data space $x \in \Xspace$. In our case, using this definition has two problems: (1) it assumes the knowledge of the prior distribution $p(x)$ over the data space; and (2) it requires the integration over the entire configuration space.
In our context of atomistic simulations, obtaining both requires exhaustively sampling the potential energy surfaces (PESes), which is undesirable.

Recently, Perez \textit{et al.} proposed the use of entropy-maximization schemes for automatic dataset generation for machine learning (ML) interatomic potentials (IPs) \citeSupp{Karabin2020EntropyMaximization2,MontesdeOcaZapiain20222}.
To bypass the problems above, the authors approximated the entropy using a classical non-parametric estimation from the literature \citeSupp{Beirlant1997Nonparametric2}. Up to a constant, that estimate is given by

\begin{equation}\label{eq:si:entropy-lanl}
    \HX_\mathrm{Perez} = \frac{1}{n} \sum_{i=1}^{n} \log \left(
        n \min_j \norm{\X_i - \X_j}
    \right),
\end{equation}

\noindent with $\X_i$ the descriptor of atom $i$, $n$ the number of descriptors in the set $\set{\X}$, and $\norm{.}$ the L$_2$ norm.
This definition is similar to that from Oganov and Valle,\citeSupp{oganov2009Howa2}

\begin{equation}\label{eq:si:entropy-oganov}
    \HX_\mathrm{Oganov} = \left\langle
    \log \left[
        1 - d\left(\X_i, \X_j \right)
    \right]
    \right\rangle,
\end{equation}

\noindent where $d$ is a custom distance function between $\X_i, \X_j$, which assumes values between 0 and 1, and $\langle.\rangle$ is the average over all structures.
One problem with these descriptions is that the nearest-neighbor distance in the descriptor space may not be a good approximation of the distribution density $p(x)$ and strongly depends on the choice of descriptor.
Furthermore, the information penalty for overlapping descriptors (i.e., $\norm{\X_i - \X_j} \ra 0$ in \eqref{eq:si:entropy-lanl}) is $\Info \ra -\infty$, which may be undesirable.
Often, when sampling PESes, oversampling certain configurations is expected, which can pose a problem to a measure of entropy that drastically penalizes any overlap between two points.
Finally, the value of entropy is unbounded, assuming any real value.
This prevents concrete analogies between atomistic datasets and information theory.

To bypass these problems, we model the distribution of data points $p(x)$ using a kernel density estimation (KDE) and use to quantify the entropy of a dataset.
This first estimate is obtained by using a normalized kernel $K_h(\X, \X_i)$ with bandwidth $h$ and averaging over all data points in a dataset $\set{\X_i}$,

\begin{equation}\label{eq:si:entropy-px}
    p(\X) = \frac{1}{n} \sum_{i=1}^{n}
        K_h(\X, \X_i).
\end{equation}

\noindent Then, as sampling the input space $\Xspace$ is undesirable when calculating the integral in Eq. \eqref{eq:si:entropy-info}, we propose an empirical estimate\citeSupp{Beirlant1997Nonparametric2} given by

\begin{equation}\label{eq:si:entropy-kde-1}
    \HX = - \frac{1}{n} \sum_{i=1}^n \log p(\X_i).
\end{equation}

\noindent This equation corresponds to the empirical entropy for a set of points $\X_i \in \set{\X}$.
Now, using Eq. \eqref{eq:si:entropy-px} to compute $\log p(\X_i)$ further simplifies this equation to

\begin{equation}\label{eq:si:entropy-kde-main}
    \HX = - \frac{1}{n} \sum_{i=1}^n
        \log \left[
            \frac{1}{n} \sum_{j=1}^n K_h (\X_i, \X_j)
        \right],
\end{equation}

\noindent To finally compute the entropy, a Gaussian kernel between descriptors can be used,

\begin{equation}\label{eq:si:entropy-kernel}
    K_h (\X_i, \X_j) = \exp \left(
        \frac{-\norm{\X_i - \X_j}^2}{2h^2}
    \right),
\end{equation}

\noindent where $\norm{.}$ is the $L_2$ norm. Along with Equation \eqref{eq:si:entropy-kde-main}, the computation of the kernel allows us to measure the information entropy of a given atomistic dataset with a single parameter $h$.

\subsection{Properties of the entropy}

The main difference between Eq. \eqref{eq:si:entropy-kde-main} and Eqs. \eqref{eq:si:entropy-lanl}, \eqref{eq:si:entropy-oganov} lies on the fact that overlapping (or completely dissimilar) points in Eq. \eqref{eq:si:entropy-kde-main} do not lead to $\Info \ra -\infty$, which is desirable when sampling potential energy surfaces.
Moreover, appropriate choice of a kernel $K_h$ abstracts away from descriptor distances and maps the entropy back to the space of probability distributions.
As a consequence, our entropy \eqref{eq:si:entropy-kde-main} exhibits the following properties:

\begin{itemize}
    \item \textbf{Bounds}: the normalization of the kernel, $0 \leq K_h (\X_i, \X_j) \leq 1$, implies that $1 \leq \sum_j K_h (\X_i, \X_j) \leq n$, so $\Info$ is bounded between 0 and $\log n$.
    \item \textbf{Minimum entropy}: $\Info = 0$ corresponds to a degenerate dataset created with multiple copies of a single $\X_i$, thus one that does not provide information about a space $\Xspace$ but only for a single point. This is exactly what one expects from $p(x) \ra \delta(x)$ in Eq. \eqref{eq:si:entropy-info}.
    \item \textbf{Maximum entropy}: $\Info = \log n$ corresponds to a dataset with zero overlap between data points, hence conveying maximal information. In information theory, this corresponds to distributions where all  outcomes are equally likely.
    \item \textbf{Entropy grows with dataset size}: Because of the $\log n$ term, datasets composed by non-overlapping data points always bring more information as the training set grows.
    \item \textbf{Entropy can decrease as new points are added}: In addition to the $\log n$ effect, if new points overlap substantially with the existing dataset, the entropy of the new dataset may be smaller than the entropy of the original dataset.
    \item \textbf{Units}: because of the reliance on the probability distributions, the entropy has units (nats) and can be used to compare datasets and descriptors. For example, for the same datasets, incomplete descriptors should have lower entropy than complete ones, as the former map two points to the same representation. For the same descriptors, richer datasets should have higher entropy than redundant ones.
\end{itemize}

These entropy properties correspond exactly to those in the field of information theory and, as a consequence of Eq. \eqref{eq:si:entropy-info}, also relate to some of those from statistical mechanics.

\subsection{Differential entropy}
\label{sec:si:dH}

In addition to the dataset entropy from Eq. \eqref{eq:si:entropy-kde-main}, one can compute the expected variation in entropy from adding a new point to the dataset even when the distances $\norm{\X - \X_i}$ are not infinite.
In information theory, this corresponds to how much information the new data brings to the dataset considering its current distribution of points.
Considering an arbitrary point in Eq. \eqref{eq:si:entropy-kde-main}, we define the differential entropy $\dH$ of adding a point $\Y$ to a dataset $\set{\X_i}_{i=1,\ldots,n}$ as

\begin{equation}\label{eq:si:entropy-kde-delta}
    \dH(\Y | \set{\X_i} ) = - \log \left[
        \sum_{i=1}^n K_h (\Y, \X_i)
    \right].
\end{equation}

\noindent This form is related to the functional derivative of the information entropy from Eq. \eqref{eq:si:entropy-info} with respect to the probability distribution $p(x)$,

\begin{equation}\label{eq:si:entropy-func-derivative}
    \frac{\dH}{\delta p(x)} = - 1 - \log p(x),
\end{equation}

\noindent thus representing the sensitivity of the entropy $\Info$ with respect to variations of its distribution $p(x)$.
In our work, we shift it for convenience by a constant $1-\log n$ (partially due to the normalization of the kernel and $p(x)$) and adopt $\dH = \log n - \log p(x)$.
Furthermore, the term ``differential entropy'' is usually employed in information theory to describe the entropy of a continuous probability distribution.
In our case, we prefer to reserve this term to the quantity given by $\dH$ and avoid using different terms for continuous or discrete probability distributions.

Equation \eqref{eq:si:entropy-kde-delta} above has interesting properties for dataset analysis and construction:

\begin{itemize}
    \item \textbf{There is no limit to ``information novelty'':}
    Contrary to $\Info$ in Eq. \eqref{eq:si:entropy-kde-main}, $\dH$ does not have an upper bound.
    If the point $\Y$ has near-zero overlap with all points $\set{\X_i}$ of the existing dataset --- and thus has maximal novelty --- then $K_h (\Y, \X_i) \rightarrow 0$ and $\dH \ra +\infty$.
    \item \textbf{Duplicating one isolated point from the training set brings zero information:} If a point $\Y$ overlaps perfectly with only one data point in $\set{\X_i}$, the sum over kernel values is one and $\dH = 0$.
    \item \textbf{Negative $\dH$ implies redundant information:} A point that overlaps with multiple points may have the summation over kernel values greater than one, leading to $\dH < 0$.
    The latter situation corresponds to points that are overrepresented in the dataset $\set{\X_i}$
    \item \textbf{Lower bound:} the differential entropy has a lower bound $- \log n \leq \dH$, where $n$ is the size of the dataset $\Xset$.
    This can only be achieved in the case where $K_h (\Y, \X_i) = 1$, and represents the scenario where all points overlap.
    The result can be interpreted as an absolute threshold for dataset redundancy.
\end{itemize}

With the properties above, it follows that the differential entropy of the points in the training set is always smaller or equal to zero, which allows us to compute uncertainties without arbitrary thresholds.

The entropy of a system can be recovered from the values of $\dH$ by

\begin{equation}\label{eq:si:dH-to-H}
    \Info (\set{\X}) = \log n - \frac{1}{n}
        \sum_{j=1}^n \dH(\X_j | \set{\X} ).
\end{equation}

\noindent Importantly, however, the differential entropy $\dH$ cannot be used to measure the entropy $\Info(\set{\X_i}_{i=1,\ldots,n+1})$ compared to $\Info(\set{\X_i}_{i=1,\ldots,n})$ when the point $\X_{n+1}$ is added to $\set{\X_i}_{i=1,\ldots,n}$.
As the new estimated probability distribution $p(x)$ changes given the knowledge of $\X_{n+1}$, the density $\frac{1}{n} \sum_j K_h (\X_i, \X_j)$ may change when the summation index is allowed to go from 1 to $n + 1$ instead of $1$ to $n$.

\subsection{Entropy in the nearest-neighbors limit}
\label{sec:si:entropy-approx}

In the limit of non-overlapping points, the sum over kernel values $K_h (\X_i, \X_j)$ from Eq. \eqref{eq:si:entropy-kde-main} can be simplified to $K_h (\X_i, \X_i) = 1$ plus the nearest neighbor value,

\begin{equation}\label{eq:si:entropy-kde-approx-1}
    \HX \approx \log{n} - \frac{1}{n} \sum_{i=1}^n
        \log \left[
            1 + \max_{j \neq i} K_h (\X_i, \X_j)
        \right],
\end{equation}

\noindent thus resembling the result from Eq. \eqref{eq:si:entropy-lanl}.
The assumption of a nearest neighbor dominance expedites the calculation of the entropy.
However, the result may not be accurate, as it requires points with small overlap in the descriptor space, an unusual assumption when dealing with PESes.
On the other hand, computing all pairwise kernels $K_h (\X_i, \X_j)$ can be expensive for a large dataset $\set{\X}$.
A good compromise is to implement the summation over the neighborhood $\mathcal{N}_k$ of $\X_i$, which contains the $k$-nearest neighbors of $\X_i$,

\begin{equation}\label{eq:si:entropy-kde-knn}
    \HX \approx - \frac{1}{n} \sum_{i=1}^n
        \log \left[
            \frac{1}{k}
            \sum_{\X_j \in \mathcal{N}_k(\X_i)}
            K_h (\X_i, \X_j)
        \right],
\end{equation}

\noindent and query the $k$-nearest neighbors with average complexity $\mathcal{O} (kd \log N)$, where $N$ is the reference dataset size.
Several approximations and nearest neighbors search methods can be employed to obtain the nearest neighbors in the feature space.
In the results shown in Sec. \ref{sec:si:approx-dH}, we used an approach based on nearest neighbors graph, which can handle dataset sizes on the order of millions, and is helpful when performing uncertainty quantification (see Sec. \ref{sec:si:methods-uq}).

The use of approximate nearest neighbors for the computation of $\dH$ is analogous to that from $\Info$,

\begin{equation}\label{eq:si:dh-kde-knn}
    \dH(\Y | \Xset) \approx - \log \left[
            \sum_{\X_j \in \mathcal{N}_k(\Y)}
            K_h (\Y, \X_j)
        \right].
\end{equation}

\noindent An immediate consequence of this approximation is that the value of $\dH$ is \textit{overestimated}, as contributions from neighbors outside of the $k$-neighborhood of each vector are neglected.
As the values of $k$ increase, $\dH$ necessarily decreases, reaching a minimum when the full dataset size is used for its computation.
Therefore, when used with the absolute threshold $\dH > 0$, the approximate $\dH$ are \textit{conservative estimates} of the uncertainty.
Some approximate nearest neighbor methods also have recall smaller than 100\%, representing the case where some of the true nearest neighbors are not recalled during the query.
Nevertheless, less accurate $\dH$ are still overestimated with respect to an ideal nearest neighbor search.
This demonstrates that, despite the approximations of truncating the expansion of $\dH$, this value can provide conservative estimates when used as an UQ metric.

\subsection{Dependence of entropy with the bandwidth}
\label{sec:si:bw}

The non-parametric estimation of the information entropy $\Info$ described in Eq. \eqref{eq:si:entropy-kde-main} requires fitting a KDE to the data distribution.
In the current work, this selection is challenging due to two issues:
(1) differences in density lead to changes in the metric space of the descriptors $\X$;
and (2) differences in entropy can vary with the choice of bandwidth.
To simplify the problem, we selected a bandwidth of 0.015 \AA$^{-1}$, adopted as constant in this work (except in Sections \ref{sec:si:thermo-bw} and \ref{sec:si:thermo}).
As described in the \methods, this corresponds roughly to the distance between two FCC environments ($k = 32$, $r_\mathrm{cut} = 5$ \AA) with an equilibrium lattice parameter of 3.58 \AA\ and another with  unit cell parameters rescaled by 1\% (see Fig. \ref{fig:si:01-x-strain}).
The use of this bandwidth to match different units of entropy is described in Sec. \ref{sec:si:thermo-bw}.

\subsection{Dataset diversity}
\label{sec:si:diversity}

As shown in Fig. \ref{fig:02-dset} of the main paper, the dataset entropy depends on how frequent each environment is sampled in the configuration space.
Therefore, entropy values can often reduce even as dataset sizes drastically increase.
To create a measure of dataset \textit{diversity} that is more robust to oversampling, we propose to express the diversity $D$ as

\begin{equation}
    D\left(\Xset\right) =
    \log \left[\sum_{i=1}^n
        \frac{1}{\sum_{j=1}^n K(\X_i, \X_j)}
    \right] =
    \log \left[
    \sum_{i=1}^n \exp\left(\dH_i\right)
    \right],
\end{equation}

\noindent where $\dH_i = \dH (\X_i | \Xset)$. This analytical form is proposed to satisfy the following properties:

\begin{itemize}
    \item \textbf{For non-overlapping datasets, $D$ recovers $\Info$:} this can be demonstrated by verifying that, in datasets where $K(\X_i, \X_j) = \delta_{ij}$, $\dH_i = 1, \forall i$ and $D (\Xset) = H (\Xset) = \log n$.
    \item \textbf{An entirely new data point increases the summation in diversity by one:} this follows from the fact that, for a new point $\X_{(n + 1)}$ that does not overlap with any of the other points $\X_i$, $\dH_{(n + 1)} = 0$.
    \item \textbf{$D$ has the same units of $\dH$,} which is determined by the base of the logarithm, and thus is nats for this work.
    \item \textbf{Repeating data points in the training set does not increase its diversity}, even if the entropy can be reduced. This follows from the summation of inverse of $p(\X_i)$, which approximately re-weights the distribution of data points based on their frequency according to other points.
\end{itemize}

Within this definition, the diversity $D$ of a dataset represents the coverage of the configuration space.
However, it does not express the same value as $\log n$, the maximum information entropy.
Whereas $\log n$ is agnostic to the coverage of the space, $D$ attempts to quantify exactly how many unique points are present in the system.
For example, a degenerate system with $\Info = 0$ also has $D = 0$ regardless of $\log n$.

\subsection{Toy examples for QUESTS}

\subsubsection{2D visualization of the entropy}

\begin{figure}[!h]
    \centering
    \includegraphics[width=0.8\linewidth]{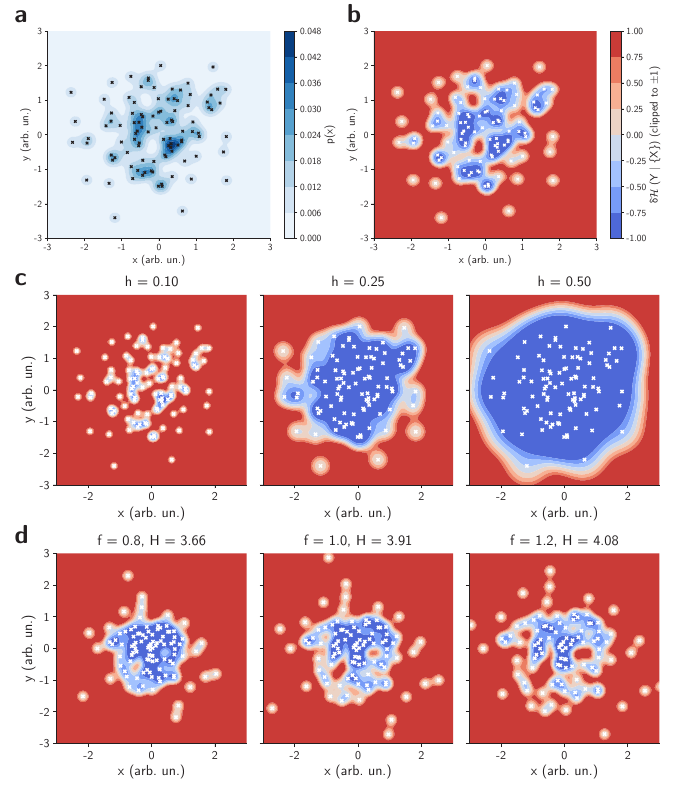}
    \caption{2D example of the KDE, bandwidth, and entropy.
    \textbf{a}, estimated $p(x)$ for a set of points (marked with x).
    \textbf{b}, the values of $p(x)$ can be mapped directly to $\dH$. This creates a common reference of $\dH > 0$ for points ``outside'' of the training set, shown here in red, and $\dH < 0$ for points ``inside'' the training set, shown in blue.
    \textbf{c}, effects of the bandwidth in estimating the probability distribution. A large bandwidth estimates the values as a single Gaussian, whereas a small bandwidth considers each point individually.
    \textbf{d}, effects of rescaling the coordinates of a distribution by a factor $f$ in the entropy $H$. Denser distributions lead to lower entropy, whereas larger spread relates to higher entropy if the bandwidth is kept constant.
    }
    \label{fig:si:01-2d-example}
\end{figure}

\begin{figure}[!h]
    \centering
    \includegraphics[width=0.8\linewidth]{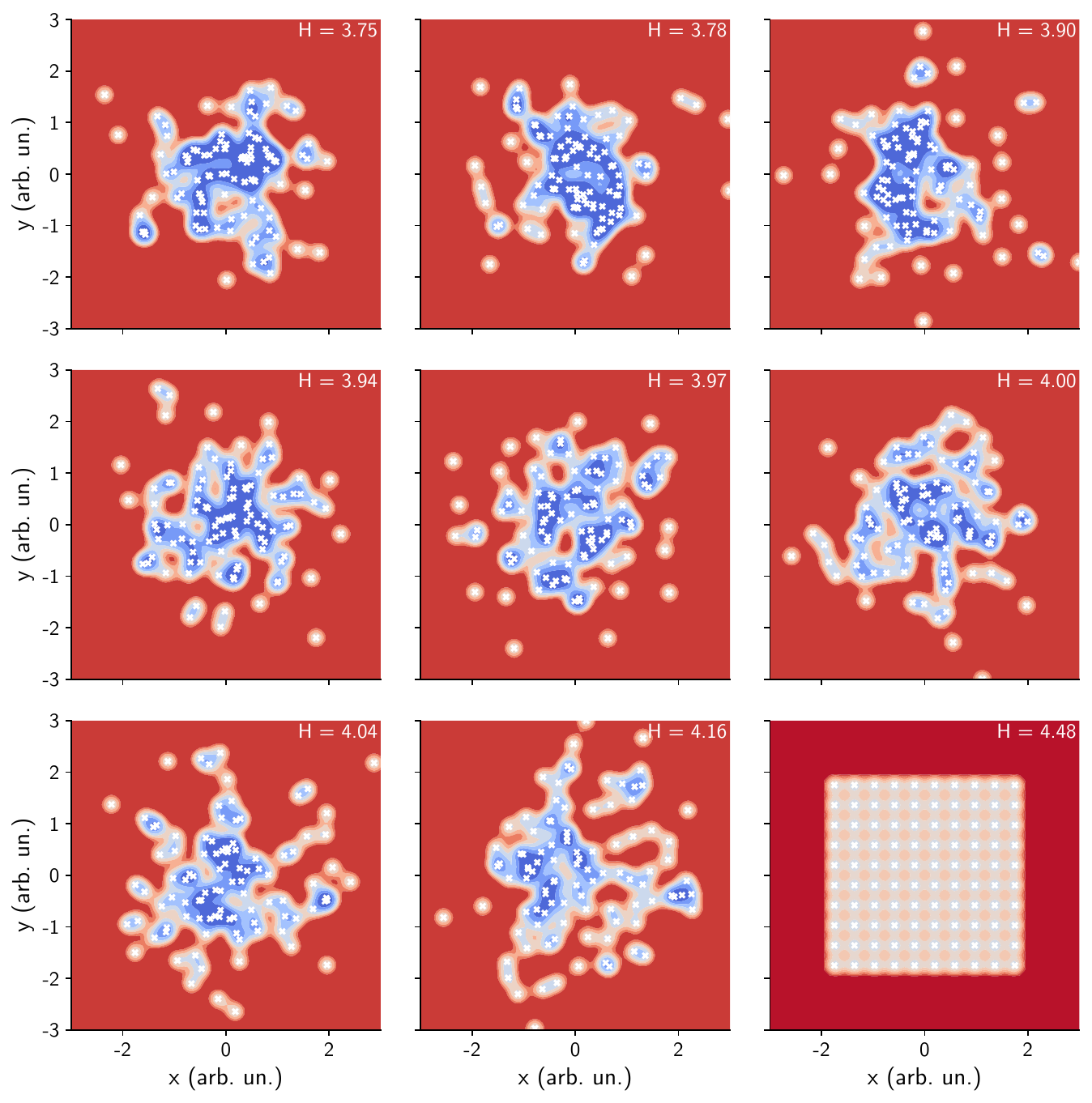}
    \caption{2D example of distributions with increasing entropy.
    The first eight distributions were generated randomly, then sorted according to their entropy.
    This provides a visual guide to interpreting values of lower entropy as more concentrated data points and higher entropy as larger spread.
    A regular occupation of the (2D) configuration space (bottom right) leads to the highest entropy among all examples.
    The color follows the same scale as \ref{fig:si:01-2d-example}b, with red points having $\dH > 0$ and blue points having $\dH < 0$.
    }
    \label{fig:si:01-2d-example-2}
\end{figure}

To visualize the concepts of entropy and distributions, we sampled 100 points in a two-dimensional space from a 2D Gaussian with mean zero and covariance matrix equal to the identity.
Then, we computed the values of $p(x)$ from a KDE and its corresponding $\dH$ for each point on the 2D grid.
Figures \ref{fig:si:01-2d-example} and \ref{fig:si:01-2d-example-2} show how the entropy $\Info$ and the differential entropy $\dH$ behave with different distributions, bandwidths, and rescaling.
If the objective was to reproduce the original Gaussian, as in a standard KDE, the choice of higher bandwidths (Fig. \ref{fig:si:01-2d-example}c) better approximates the actual distribution.
While this example is more difficult to visualize in a high-dimensional space of atomistic environments,  the distribution plots illustrate the analogous result that would happen to them.

\subsubsection{Visualization and distance for the atomistic representation}

\begin{figure}[!h]
    \centering
    \includegraphics[width=0.5\linewidth]{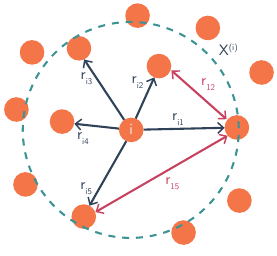}
    \caption{
    Visualization of the distances used to create the $\X_1$ and $\X_2$ representation.
    }
    \label{fig:si:01-descriptors}
\end{figure}

\begin{figure}[!h]
    \centering
    \includegraphics[width=\linewidth]{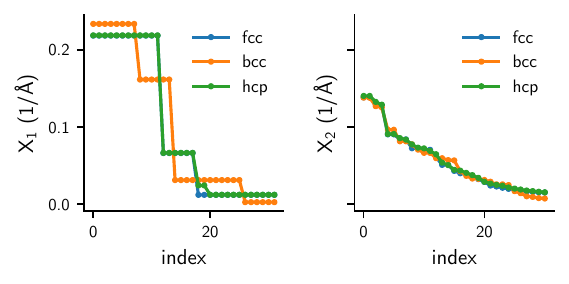}
    \caption{
    Visualization of the $\X_1$ and $\X_2$ representation for FCC, BCC, and HCP structures.
    The small differences between FCC and HCP can be seen only at neighbors further away from the origin.
    }
    \label{fig:si:01-x-descriptors}
\end{figure}

\begin{figure}[!h]
    \centering
    \includegraphics[width=\linewidth]{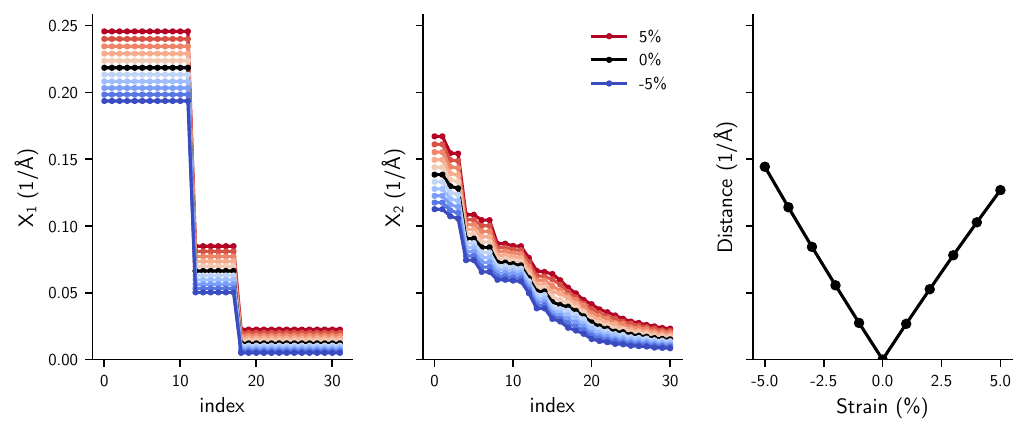}
    \caption{
    Behavior of the $\X_1$ (left) and $\X_2$ (middle) representation for an FCC structure under strain between -5\% (expansion, blue) and 5\% (compression, red).
    The Euclidean distance between the strained and reference structure is shown on the right.
    Within this range of uniform strains and this structure, the distance varies linearly.
    }
    \label{fig:si:01-x-strain}
\end{figure}

The representation proposed in this work was created on a per-environment basis, with radial distances and cross distances, as explained in Section \ref{sec:repr} above and shown in Fig. \ref{fig:si:01-descriptors}.
The representation can be visualized in a single plot and used to differentiate between standard crystal structures, such as BCC, FCC, and HCP (Fig. \ref{fig:si:01-x-descriptors}).
This descriptor can also be used upon modification of the original structure, such as strain.
In Fig. \ref{fig:si:01-x-strain}, an FCC structure is strained between -5\% and 5\%, and the representation is visualized according to the applied strain.
Interestingly, the distance between the descriptors and the applied strain varies near-linearly within this range.

\subsubsection{Information entropy upon denoising}

\begin{figure}[!h]
    \centering
    \includegraphics[width=0.5\linewidth]{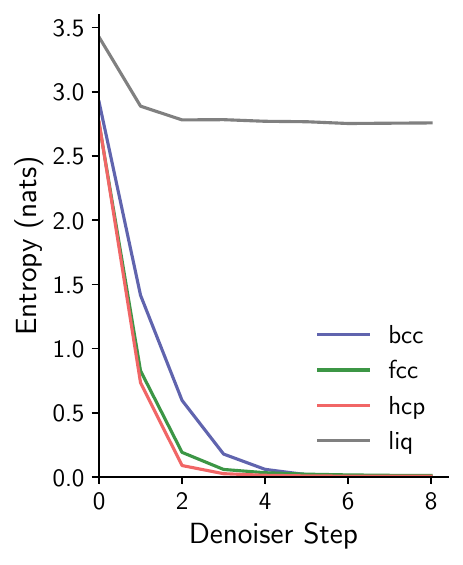}
    \caption{
    Information entropy of four phases of copper (FCC, HCP, BCC, and liquid) for the denoised trajectories from Hsu et al.
    }
    \label{fig:si:01-x-denoiser}
\end{figure}

To exemplify how the entropy $\Info$ and the descriptors can be used to quantify information within atomistic systems, we analyzed trajectories with decreasing diversity of atomic environments from Hsu et al.\citeSupp{hsu2023score2}
Because the deviations of the atoms from their ideal lattice sites were removed with a denoising method to enable phase classification, we expect the values of $\Info$ to decrease accordingly.
To validate this intuition for $\Info$, we computed the information entropy of four denoised phases of copper, as shown in Fig \ref{fig:si:01-x-denoiser}.
As vibrational motion is removed from the system, the values of $\Info$ for the crystalline phases FCC, BCC, and HCP decrease until reaching zero.\citeSupp{hsu2023score2}
On the other hand, the liquid phase cannot be fully denoised, and the residual disorder is manifested in a higher information entropy value.
This example illustratse a connection between configurational degrees of freedom and information $\Info$ of atomistic structures.

\subsection{Extended discussion on learning curves of the rMD17 dataset}
\label{sec:si:rmd17}

\begin{figure}[!h]
    \centering
    \includegraphics[width=0.7\linewidth]{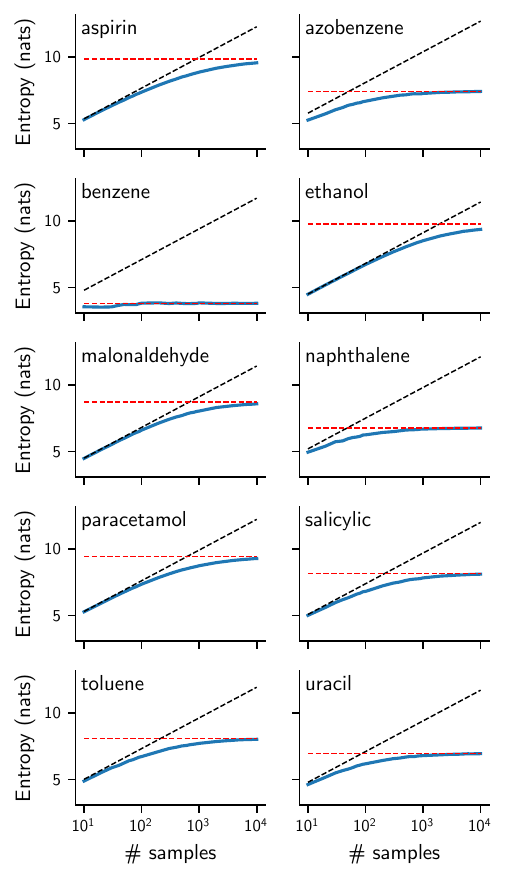}
    \caption{
    Entropies for all rMD17 molecules as a function of training set size.
    The black dashed line is the behavior of $\log n$ considering the number of environments per molecule.
    The red line is the asymptote for the entropy.
    }
    \label{fig:si:04-rmd17-all-entropies}
\end{figure}

\begin{figure}[!h]
    \centering
    \includegraphics[width=0.7\linewidth]{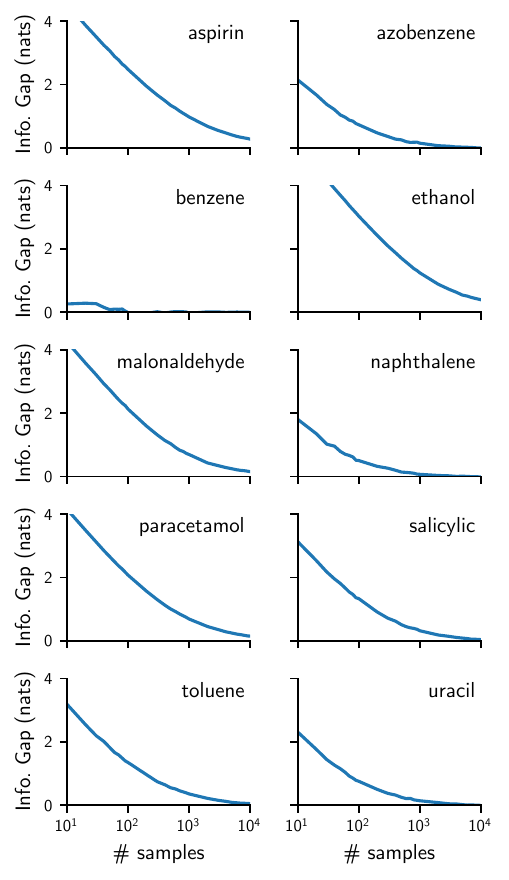}
    \caption{
    Information gap for all rMD17 molecules as a function of training set size.
    The gap is defined as the asymptotic value of the information entropy minus the entropy value at a given number of samples.
    These curves show that, at a typical constant number of samples, the information gap varies substantially across molecules.
    }
    \label{fig:si:04-rmd17-info-gap}
\end{figure}

\begin{figure}[!h]
    \centering
    \includegraphics[width=0.7\linewidth]{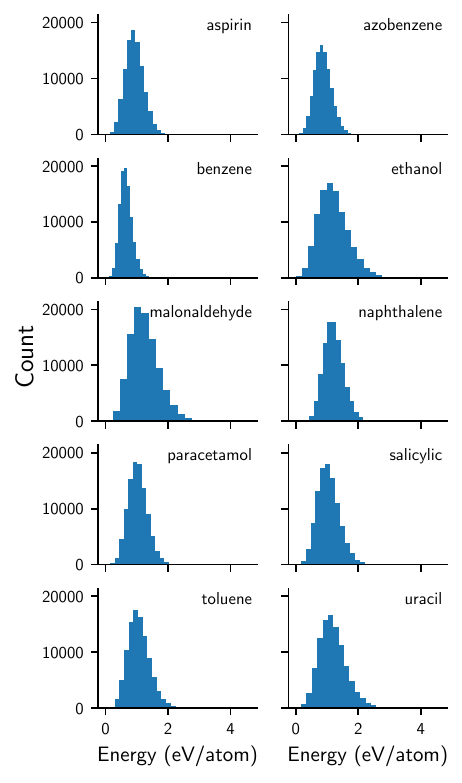}
    \caption{
    Distribution of energies for the original rMD17 dataset.
    Ethanol and malonaldehyde have larger standard deviations and longer tails towards higher energies.
    }
    \label{fig:si:04-rmd17-energy-hist}
\end{figure}

\begin{figure}[!h]
    \centering
    \includegraphics[width=\linewidth]{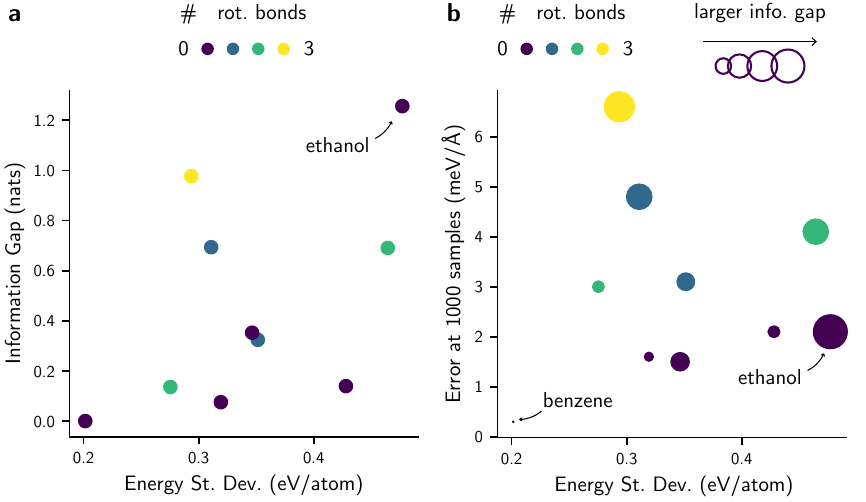}
    \caption{
    Correlation between the force errors from a MACE model, the standard deviation of the distribution of energies, and the information gap for each molecule (represented with marker sizes).
    For the systems with zero rotatable bonds, the error is higher for wider distributions.
    }
    \label{fig:si:04-rmd17-energy-std}
\end{figure}

Figure \ref{fig:02-dset}a of the main text discusses the trends in learning curves for different molecules in the rMD17 dataset, and Fig. \ref{fig:si:04-rmd17-all-entropies} shows the results of all molecules.
Within the discussion, ethanol can be perceived as an outlier for this trend.
Despite being much smaller than the other molecules, its information entropy takes a long time to reach a maximum compared to its counterparts, thus increasing its information gap (Fig. \ref{fig:si:04-rmd17-info-gap}), which is unexpected at first.
To explain this result, we notice that the distribution of energies for the rMD17 dataset varies according to the molecule (Fig. \ref{fig:si:04-rmd17-energy-hist}).
Molecules such as ethanol and malonaldehyde, despite small, have broader distributions compared to their counterparts, which correlates positively with higher information gaps (Fig. \ref{fig:si:04-rmd17-energy-std}a).
Thus, if we assume that energy distributions correlate with the accessible phase space on a per-system basis, then the information gap correctly captures this effect for the molecules, including ethanol, explaining this counterintuitive outlier.

\subsection{Cluster size distribution and classical nucleation theory}
\label{sec:si:cnt-clusters}

Beyond the critical nuclei described in the main text, we verify that the distribution of cluster sizes in the melt can also be predicted using our approach.
Figure \ref{fig:si:04-solid-cna} shows that, for all pre-nucleation snapshots, the cluster sizes follow approximately a power law.
An analytical expression derived from the CNT (Sec. \ref{sec:si:methods}), when fit to the data, also matches the data distribution, with a predicted surface energy of about 0.104 J/m$^2$.
While this value underestimates the experimental range of 0.177--0.221 J/m$^2$ (described in the main text) it is still remarkably close to the overall data considering the approximations of the cluster definition, surface-to-volume ratios, and other factors not accounted for in our approach.
Obtaining this fit allows us to estimate system properties relevant for CNT without relying on direct measurements of the surface areas or Gibbs energies, thus providing useful insights on the physical phenomenon and further demonstrating the usefulness of our method.

\subsection{Qualitative parallels between information entropy and thermodynamics}
\label{sec:si:thermo-sec}

\subsubsection{Information entropy and heat capacity}

\begin{figure}[!h]
    \centering
    \includegraphics[width=0.7\linewidth]{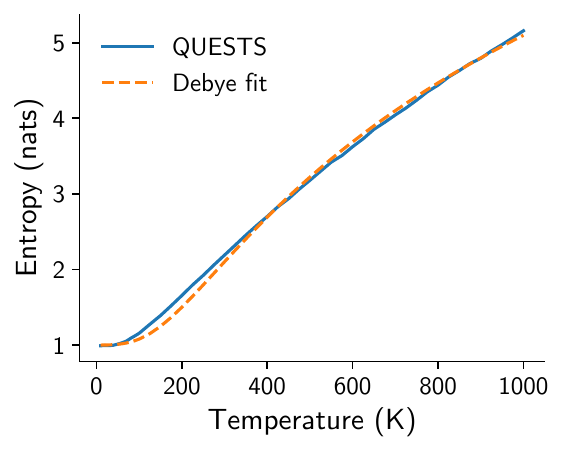}
    \caption{Information entropy of particles interacting via harmonic bonds under a quantum thermal bath (QTB). At zero temperature, the entropy does not go to zero to simulate the effects of the zero-point energy. A fitted Debye model is shown with a dashed orange line.}
    \label{fig:si:01-debye}
\end{figure}

One of the simplest toy models for entropy is the Debye's model, which considers atoms interacting via harmonic potentials as a model for phonons and heat capacity.
To obtain classical MD trajectories that match the physics from the Debye model (and the zero-point energy in quantum harmonic oscillators), we used the quantum thermal bath (QTB) implemented in LAMMPS.\citeSupp{dammak2009quantum2}
We simulated a $10 \times 10 \times 10$ box of particles with the FCC structure, unit cell parameter of 3.645 \AA, and mass of 62.5 g/mol.
The bond terms are determined by the spring constant $k = 1.0$ eV/\AA$^2$ and an equilibrium distance of $2.5775$ \AA.
Bonds are created for particles that are between 2.0 and 3.0 \AA\ apart.
Then, the simulation is performed using a QTB at constant temperature, varying from 10 to 1000 K, $f_\mathrm{max} = 120$ ps$^{-1}$, $N_f = 100$, constant volume.
The simulation was performed with an equilibration run of 300 ps and a production run of 100 ps using a timestep of 2 fs.
The results are shown in Fig. \ref{fig:si:01-debye}.
Although the entropy was obtained with a constant, low value of bandwidth (0.015 \AA$^{-1}$), the entropy of a fitted Debye model matches closely that from the extracted simulations.
Importantly, the entropy does not approach zero at 0 K due to the residual motion from the simulations that mimic the behavior of the zero-point energy.

\subsubsection{Calibrating the information entropy to configuration entropy differences with the bandwidth}
\label{sec:si:thermo-bw}

Because lower densities (higher atomic volumes) lead to lower distances in the descriptor spaces, calibrating the information and thermodynamic entropies may be used with a variable bandwidth that decreases with increasing atomic volume,

\begin{equation}
    h(V) = a \exp\left(-b V^2\right) + c,
\end{equation}

\noindent where $a$, $b$, and $c$ are unknown parameters.
To estimate these parameters in a self-consistent way, we first performed simulations for the copper Einstein crystal at the NVT ensemble using the spring constant of 34.148 eV/\AA$^2$ and for volumes from 6 to 50 \AA$^3$/atom.
Though this method may vary slightly with the choice of spring constant and temperatures, we observed that the selected bandwidth was transferable across many systems without refitting, as discussed in Sec. \ref{sec:si:thermo}.
The \texttt{fix ti/spring} command in LAMMPS was used with a value of $\lambda$ that ensures that only the harmonic oscillator is considered in the simulation.
Then, for each volume, the entropy of the system is computed for a range of bandwidths, varying from 0.010 to 0.090 \AA$^{-1}$.
As the entropy of the Einstein crystal is independent of the volume, we estimate the values of bandwidth that would keep the entropy reasonably constant across the range of volumes.
Figure \ref{fig:si:01-bandwidth} shows the results of this investigation, and the fitted bandwidth prediction that rescales the (arbitrary) information entropy to the thermodynamically relevant units $k_B$/atom.
This approach provides a systematic rescaling of the bandwidths, but unfortunately does not guarantee that this scaling is universal.
Though all examples demonstrated in Sections \ref{sec:si:lindemann} and \ref{sec:si:thermo} use this same scaling of the bandwidth, extension of this analysis beyond the toy examples discussed here will be an object of future investigation.

\subsubsection{Information entropy and Lindemann's melting criterion}
\label{sec:si:lindemann}

\begin{figure}[!h]
    \centering
    \includegraphics[width=0.8\linewidth]{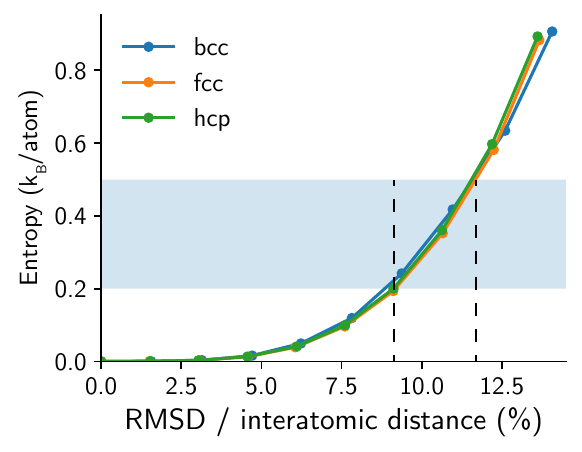}
    \caption{
    Root mean square deviation (RMSD) of atoms in FCC, BCC, and HCP structures, and their corresponding entropies calculated with QUESTS.
    The shaded area represents typical solid entropies prior to melting, and intersects the computed curves between 10--12.5\% RMSD/interatomic distances, thus reproducing the Lindemann melting rule.
    }
    \label{fig:si:01-lindemann}
\end{figure}

The Lindemann melting criterion is a well-known estimate for the melting point of materials.\citeSupp{lindemann1910melting2}
According to this estimate, melting often happens when the ratio between the root mean square displacement (RMSD) of atoms with respect to their ideal lattice positions and the ideal interatomic distances approaches a constant factor, often around 0.10 for several metals.
To verify if our method could reproduce these results, we gradually added noise to the positions of prototypical FCC, BCC, and HCP crystal structures.
To obtain a statistically meaningful result, we employed a $25 \times 25 \times 25$ supercell for each of the structures, thus creating structures with 15,625 atoms.
Then, for each level of noise, we computed the RMSD with respect to the ideal lattice sites, and the entropy of the noisy configuration.
When a bandwidth of 0.057 \AA$^{-1}$ is used for the chosen volumes (Fig. \ref{fig:si:01-bandwidth}), the resulting entropy is shown in Fig. \ref{fig:si:01-lindemann}.
The results show that the entropy increases rapidly with the RMSD, and reaches values around 0.2 to 0.5 $k_B$ with a normalized RMSD between 0.1 and 0.125.
As typical entropies of solids prior to melting are around this range of 0.2 to 0.5 $k_B$, considering the entropy of a liquid around 1.3 $k_B$ and melting entropies between 0.8 and 1.1 $k_B$, this result reproduces the intuition behind Lindemann's melting rule based on the entropy values.
While many other factors are responsible for melting and the Lindemann criterion is a rough approximation, this toy example shows that the addition of noise to the system leads to entropy values compatible with expected ranges.

\subsubsection{Comparing information theory and configuration entropy in toy systems}
\label{sec:si:thermo}

\begin{figure}[!ht]
    \centering
    \includegraphics[width=\linewidth]{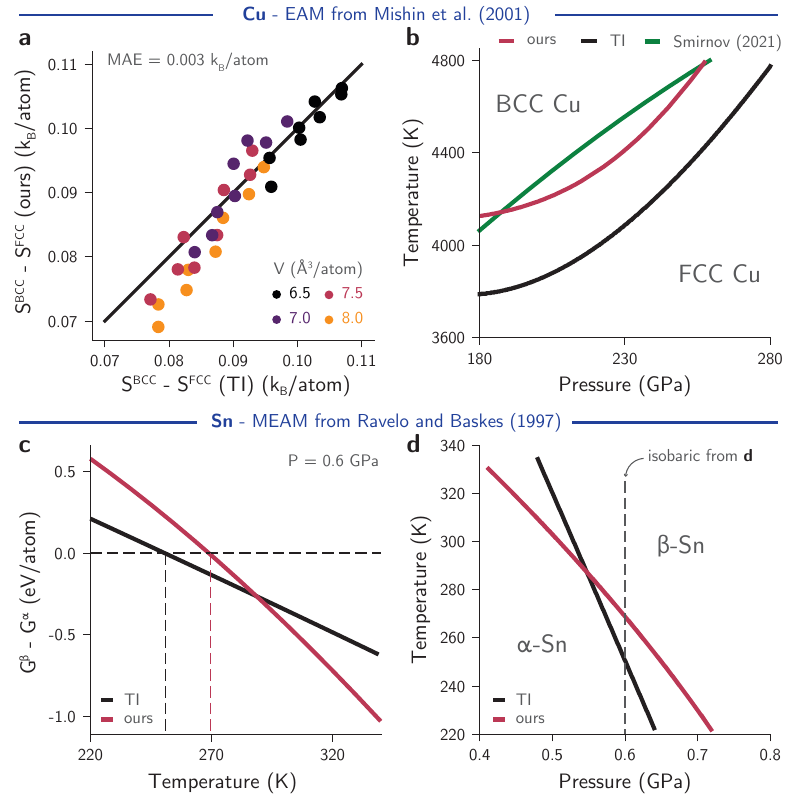}
    \caption{
    \textbf{a}, Entropy differences between BCC and FCC Cu at different temperatures and densities, as obtained by thermodynamic integration (TI) and our method, are nearly identical.
    Higher atomic volumes are shown with brighter colors.
    Different points with the same color correspond to different temperatures at the same volume.
    \textbf{b}, Phase boundaries of Cu computed using  our method (red) and from TI (black) using a force field are similar in shape and ranges.
    A reference phase boundary from the literature, computed using DFT and a quasiharmonic approximation from Smirnov, is shown in green.
    \textbf{c}, Differences in Gibbs free energy between $\alpha$ and $\beta$ phases of Sn at 0.6 GPa using our method (red) and TI (black).
    Despite the different approaches to compute $G$, the results are consistent in values and correctly predict a phase transformation around the same temperature ranges.
    \textbf{d}, The phase boundaries between $\alpha$-Sn and $\beta$-Sn computed using our method (red) and TI (black) show good agreement across a range of pressures and temperatures.
    }
    \label{fig:si:02-thermo}
\end{figure}

Using the information entropy defined in Eq. \eqref{eq:entropy-main} of the main text, we verified whether non-parametric descriptor distributions derived from atomistic simulations can be used to predict the configuration component of thermodynamic entropy differences, i.e., the entropy due to uncertainty in positions, but not momenta nor composition.
As a reference, we compared our analysis to entropy differences obtained from thermodynamic integration (TI) at constant temperature and volume/pressure.
In particular, we computed phase diagrams for two well-known systems using classical simulations: the BCC-FCC phase boundary of Cu under high pressures and temperatures ($180 \leq P \leq 280$ GPa, $3600 \leq T \leq 4800$ K), and the $\alpha$ to $\beta$ phase transformation of tin around 286 K.
As entropy differences in solid-solid phase transformations tend to be small, often smaller than one Boltzmann constant $k_B$, obtaining exact entropies is essential to produce accurate phase diagrams from simulations.
We started by performing MD simulations of Cu at low atomic volumes (6.5--8.0 \AA$^3$/atom) in the NVT ensemble using a classical IP based on the embedded atom method (EAM) from Mishin et al.\citeSupp{mishin2001structural2}
For each temperature, volume, and phase, we obtained the Helmholtz free energy $F$ within the TI method and calculated the entropy by taking the derivative of the free energy with respect to the temperature (see Sec. \ref{sec:si:methods}).
Then, we computed the reference entropy difference between the BCC and FCC phases at each volume and temperature.
To compare our information theoretic method against these TI-derived entropies, we performed MD simulations at the same (V, T) pairs, but without the coupled Hamiltonian used for the reference free energy;
instead, we use Eq. \eqref{eq:entropy-main} to analyze the information entropy of the descriptor distributions.
At a bandwidth of approximately 0.082 \AA$^{-1}$ (see Fig. \ref{fig:si:01-bandwidth}), the differences of information entropy approximate well those obtained with TI, with a mean absolute error (MAE) of 0.003 $k_B$/atom (Fig. \ref{fig:si:02-thermo}b).
Systematic deviations from the TI entropies are found as the volume increases, which could be an artifact of the selected bandwidth or functional form of the descriptors.
Nevertheless, despite the approximations from the descriptors and KDE, we successfully recovered not only trends in thermodynamic values, but also the exact values of entropy differences for the BCC and FCC Cu.
Using the energy values from the same simulations, we compared the phase boundary from both methods by mapping the Helmholtz free energy space F(V, T) into a Gibbs G(P, T) phase diagram (\methods).
The BCC-FCC phase boundaries for Cu within the ranges of 180--280 GPa and 3600--4800 K are similar in shape and values despite the impact of small entropy errors in phase boundary shifts (Fig. \ref{fig:si:02-thermo}c).
Nevertheless, the phase boundary computed with the EAM potential and our QUESTS method is close to a phase boundary from the literature,\citeSupp{smirnov2021Relative2} which was obtained using density functional theory (DFT) calculations and the quasi-harmonic approximation.
Although an ideal free energy method would recover the exact boundary obtained from the TI, this comparison suggests that our method is within reasonable deviation from the original results.

To demonstrate that entropy differences can be computed beyond constant volume assumptions, we analyzed the phase transformation between the $\alpha$ and $\beta$ phases of tin using the modified EAM (MEAM) potential from Ravelo and Baskes.\citeSupp{ravelo1997Equilibrium2}
In this transformation, the density undergoes a change of approximately 20\% from $\alpha$- to $\beta$-Sn.
First, we obtain the free energies with TI by mapping from the NVT to NPT space to ensure the consistency of the calculation at different values of $\lambda$ (see \methods).
On the other hand, our QUESTS approach allows computing information entropies directly from NPT simulations for each phase.
From these results, we compute the free energy differences at each (P, T) as $\Delta G = \Delta U - T \Delta S + P \Delta V$, where $U$ and $V$ are obtained from the average energies and volumes during the simulations.
Figure \ref{fig:si:02-thermo}d shows that the free energy differences between our method and TI at constant pressure of 0.6 GPa are in reasonable agreement.
Small errors in entropy differences in our method lead to a larger derivative of the free energy curve and overestimate the transition temperature by about 10\%.
Across a range of pressures and temperatures, the agreement between our method and TI is shown on the phase diagram of Fig. \ref{fig:si:02-thermo}e.
Although differences in transition temperatures suggest that the accuracy of our method can be further improved, this surprising agreement between descriptor distributions, information entropy, and statistical mechanics can spark future investigations on their connection.

\subsection{Approximate computation of entropy and nearest neighbors}
\label{sec:si:approx-dH}

\begin{figure}[!ht]
    \centering
    \includegraphics[width=0.8\linewidth]{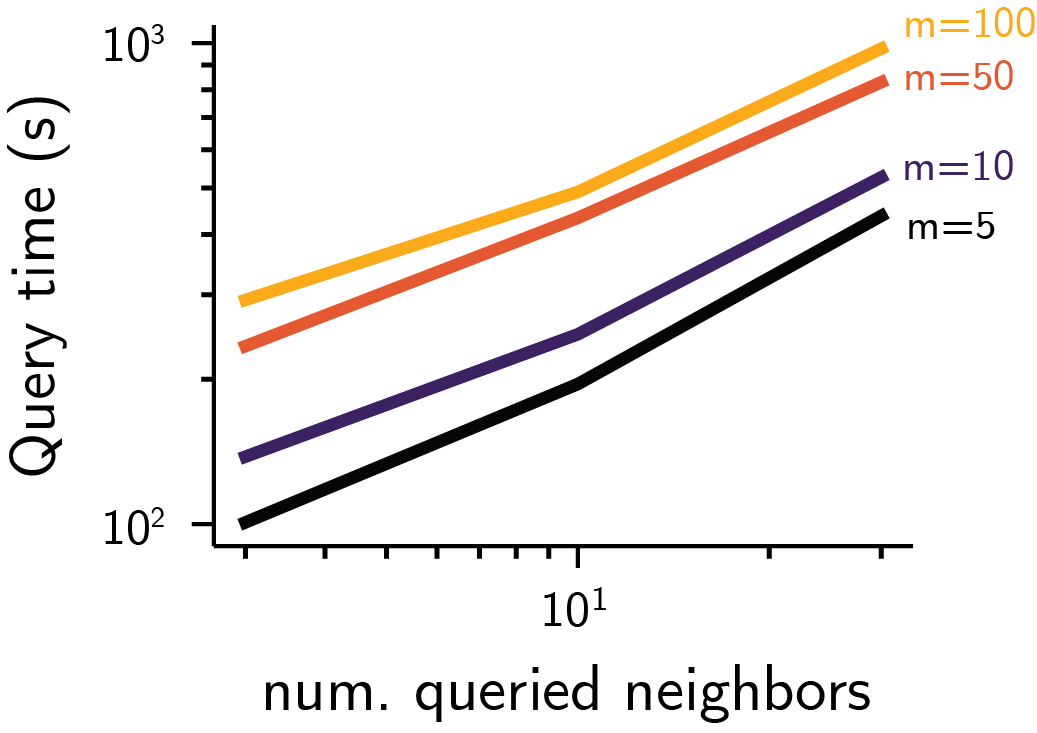}
    \caption{
    Computational performance of the approximate nearest neighbors search.
    At the low-resource side ($N = 3$ queried neighbors per environment, index constructed with $m = 5$ neighbors), the values of $\dH$ for all 32.5M atoms are evaluated in about 100 seconds when performed in a single node with 56 threads.
    For the SNAP dataset, the true $\dH$ for all environments is computed in about 255 s (wall-time) with the same hardware and parallelization settings.
    }
    \label{fig:si:05-Ta-query}
\end{figure}

At larger scales, one drawback of computing entropy values is the necessity of computing kernel matrices between each test point and the entire training set.
As the number of test points $n_Y$ and training examples $n_X$ grow, the cost of computing such matrices increases with $\mathcal{O}(n_X n_Y)$.
To verify if this is a problem in a large atomistic model, we approximate the values of $\dH$ by truncating the summation in Eq. \eqref{eq:dH} and using an approximate nearest neighbors approach (see \supptext, Section \ref{sec:si:entropy-approx}), which decreases the complexity to $\mathcal{O}(n_Y N \log n_X)$, with $N$ the number of neighbors in the descriptor space.
As computing $\dH$ for each point $\Y$ is an embarrassingly parallel task, the search can be distributed over different processes or threads to expedite the computation of this differential entropy.
Figure \ref{fig:si:05-Ta-query} shows the total query times for the 32.5M environments of tantalum relative to the SNAP training set (4224 environments) as a function of approximate nearest neighbors parameters and parallelized over 56 threads.
As the index is constructed to increase the accuracy of the approach (higher values of $m$, see Sec. \ref{sec:si:methods}), larger query times are obtained, with the slowest time obtained when an index with $m = 100$ is created and $k = 30$ neighbors are queried for each of the 32.5M test environments.
In that case, the computation of $\dH$ used a wall time of  1000 seconds when parallelized on 56 threads on 56 Intel Xeon CLX-8276L CPUs from the Ruby supercomputer.
On the other hand, the fastest set of parameters ($m = 5$, $k = 3$, 56 threads) spent 100 seconds in the same hardware.
As a reference, computing the exact $\dH$ values for the 32.5M atom system with respect to the SNAP dataset (4224 environments) takes a walltime of about 255 seconds using the same hardware and parallelization settings.
While the approximate $\dH$ has better scaling for larger reference datasets and is not critical for the SNAP dataset, performing the nearest neighbor search adds additional time constants compared to the brute-force exact calculation of the true $\dH$.
While the timings can further be improved with additional parallelization, code optimization, or use of GPU architectures, our results already demonstrate that the computation of the differential entropy, either in approximate or complete way, is accessible even for systems with a large number of environments.
The results also illustrate the theoretical understanding (Sec. \ref{sec:si:entropy-approx}) that the approximate values of $\dH$ are overestimated compared to the actual $\dH$ values, as shown in Fig. \ref{fig:si:04-approx-dH}.

\section{Supplementary Methods}
\label{sec:si:methods}

These Supplementary Methods describe in more detail the additional results and calculations in the \supptext.

\subsection*{Molecular dynamics simulations}

\noindent All MD simulations were performed using the Large-scale Atomic/Molecular Massively Parallel Simulator (LAMMPS) software\citeSupp{Thompson2022LAMMPS2} (v. 2/Aug./2023).
All simulations were performed using a 1 fs time step, except when stated otherwise.

\noindent\textbf{Thermodynamic Integration:} free energies of solids were computed by assuming a potential energy $U_\lambda$ that couples a reference system with potential energy $U_\mathrm{ref}$ and the interacting one $U_\mathrm{IS}$ such that

\begin{equation*}
    U_\lambda = \lambda^2 U_\mathrm{IS} + (1 - \lambda^2) U_\mathrm{ref},
\end{equation*}

\noindent where the quadratic term $\lambda^2$ reduces the impact of sampling the space of $(N, V, T, \lambda)$ with a uniform grid in $\lambda$, and thus creates a denser sampling around $\lambda = 0$ or $\lambda = 1$ which mitigates numerical integration errors.
The Helmholtz free energy $F$ of the interacting system is obtained first taking the derivative of the free energy of the system corresponding to $U_\lambda$ with respect to $\lambda$,

\begin{equation*}
    \left(\frac{dF_\lambda}{d\lambda}\right)_{N,V,T} =
    \left\langle
    \frac{\partial U_\lambda}{\partial \lambda}
    \right\rangle_\lambda,
\end{equation*}

\noindent where $U$ is the energy of the system. Integrating the expression above in $\lambda$, we obtain

\begin{equation*}
    F_\mathrm{IS} = F_\mathrm{ref}
    + \int_0^1
    2 \lambda
    \left\langle
    U_\mathrm{IS} - U_\mathrm{ref}
    \right\rangle_\lambda d\lambda,
\end{equation*}

\noindent where $F_\mathrm{ref}$ is known for any given temperature and volume.
We adopted the Einstein crystal as the reference, and modified  the \texttt{fix ti/spring}\citeSupp{freitas2016nonequilibrium2} in LAMMPS to obtain energies for each $(V, T, \lambda)$ without using a switching function.
Using this, we performed different simulations for each point of the grid, thus ensuring stricter convergence of the average energy differences $U_\mathrm{IS} - U_\mathrm{ref}$ for each $\lambda$.
We used a uniform grid with a spacing of 0.02 for $\lambda$, leading to 51 data points for each phase and $(V, T)$.
Numerical integration was performed using the function from the QUADPACK library\citeSupp{piessens2012quadpack2} interfaced by SciPy\citeSupp{2020SciPy2} (v. 1.11.1).

\noindent\textbf{Entropy from TI:} given the free energy computed using the TI method, the entropy by taking the derivative of the Helmholtz free energy with respect to the temperature,

\begin{equation*}
    S = -\left(
    \frac{\partial F}{\partial T}
    \right)_{N,V}.
\end{equation*}

\noindent As the free energy is not computed for an infinitely dense grid of $(V, T)$ values, numerical derivatives can lead to inaccurate values of entropy.
To mitigate this problem, we fit a quadratic 2D polynomial to the free energies as a function of the independent variables $(V, T)$.
The fit is performed using the Lasso method ($L_1$ regularization) for all polynomial features up to degree 2 using the scikit-learn\citeSupp{sklearn2} (v. 1.3.0) library, with $\alpha = 10^{-4}$ and a maximum of $10^6$ iterations.
Then, with the interpolated values of free energy, we obtain the entropy by taking the numerical derivatives of $F$ with a fine grid of temperatures at each value of volume.

\noindent\textbf{Phase diagrams from TI:} given the convenience of using the NVT ensemble when performing thermodynamic integration calculations, we constructed P-T phase diagrams by first obtaining free energies in the $(N, V, T)$ space.
Then, using the value of average pressure for each volume, we map each point $(P, T)$ into a volume $V$, and the resulting $(V, T)$ into a free energy $F$.
With these variables, we compute the Gibbs free energy as $G(P, T) = F(V(P, T), T) + P \times V(P, T)$.
The functions $(V, T) \rightarrow F$ and $(P, T) \rightarrow V$ are performed as described before, thus using a two-dimensional polynomial regressor with degree 2 and $L_1$ regularization.
We observed that direct mappings $(P, T) \rightarrow F$ led to numerical inconsistencies that drastically affected the outcomes of the phase diagram, especially given the small entropy differences between the phases.
On the other hand, the step-wise mapping was found to be more numerically stable.

\noindent\textbf{FCC-BCC Cu phase transition at high pressure:}
the phase boundary between the FCC and BCC phases of copper was simulated using the EAM potential from Mishin \textit{et al.}\citeSupp{mishin2001structural2}
The phases were simulated at four volumes: 6.5, 7.0, 7.5, and 8.0 \AA$^3$/atom, which correspond to the range of high pressures shown in Fig. \ref{fig:si:02-thermo}b.
All calculations were performed with $20 \times 20 \times 20$ supercells, leading to an FCC cell with 32,000 atoms and a BCC cell with 16,000 atoms.
Simulations were performed at 9 temperatures between 3000 and 5000 K separated by 250 K, and 51 values of $\lambda$.
The MD simulation was performed at the NVT ensemble with the Langevin thermostat implemented in LAMMPS\citeSupp{schneider1978molecular2} and a damping constant of 0.5 ps.
The simulation was equilibrated for 100 ps before a 1 ns-long production run.
During the production run, the pressure, energy, and the coupled energy $U_\mathrm{IC} - U_\mathrm{ref}$ were averaged for every time step, and later printed for post-processing in the TI approach.
A spring constant of 34.148 eV/\AA$^2$ was used to attach the Cu atoms to their ideal lattice sites, thus modeling the Einstein crystal.

MD trajectories for entropy calculations using our QUESTS method were performed in the NVT ensemble using the same temperatures and volumes as the TI method.
Simulations used the same cell sizes as the TI, but had 100 ps-long production runs.
Snapshots were saved every 2.5 ps.
Entropy values were obtained by randomly sampling 200,000 environments of the saved trajectory with a variable bandwidth determined by their volume.

\noindent\textbf{$\alpha-$ to $\beta-$Sn phase transition:}
the phase boundary between the $\alpha$ and $\beta$ phases of tin was simulated using the MEAM potential from Ravelo and Baskes\citeSupp{ravelo1997Equilibrium2}.
The equilibrium lattice parameters for these structures were found to be $a_\alpha = 6.483$ \AA, $a_\beta = 5.830$ \AA, and $c_\beta = 3.183$ \AA.
All calculations were performed with a $12 \times 12 \times 12$ supercell for $\alpha$ and $12 \times 12 \times 24$ for $\beta$, leading to a cell with 13,824 atoms each.
For the TI, simulations were performed at three different volumes, corresponding to 98\%, 100\%, and 102\% of the equilibrium volumes of each phase, 7 temperature values between 200 and 350 K spaced by 25 K, and 51 values of $\lambda$.
The MD simulation was performed at the NVT ensemble with the Langevin thermostat implemented in LAMMPS\citeSupp{schneider1978molecular2} and a damping constant of 0.5 ps.
The simulation was equilibrated for 40 ps before a 500 ps-long production run.
During the production run, the pressure, energy, and the coupled energy $U_\mathrm{IC} - U_\mathrm{ref}$ was averaged for every time step, and later printed for post-processing in the TI approach.
A spring constant of 2.0 eV/\AA$^2$ was used to attach the Sn atoms to their ideal lattice sites, thus obtaining an ideal Einstein crystal as reference system.

\begin{figure}[!h]
    \centering
    \includegraphics[width=0.7\linewidth]{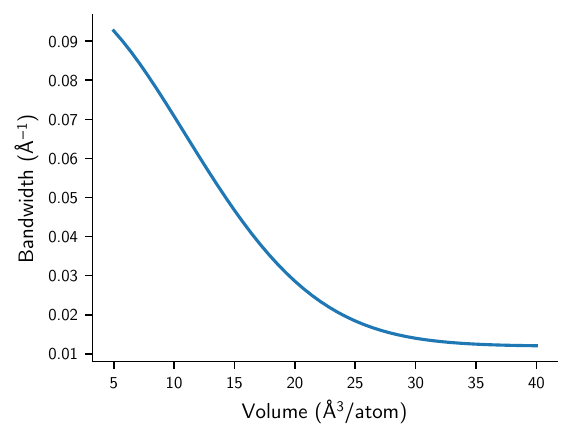}
    \caption{Proposed dependence of the kernel bandwidth with the volume. The bandwidth saturates at high volumes to ensure that residual information is captured from the data despite the non-thermodynamic behavior.}
    \label{fig:si:01-bandwidth}
\end{figure}

Entropy calculations using our QUESTS method were performed in the NPT ensemble at 1 bar and same temperatures as the TI method.
Simulations used the same cell sizes as the TI, but had 200 ps-long production runs, with snapshots saved every 10 ps.
Entropy values were obtained by randomly sampling 100,000 environments of the saved trajectory with a constant bandwidth of 0.038 \AA$^{-1}$, which corresponds to the bandwidth for the average of the volumes between the $\alpha$ and $\beta$ phases (Fig. \ref{fig:si:01-bandwidth}).

\subsection*{Classical nucleation theory analysis}

\noindent\textbf{Cluster size distribution:} within the CNT, the expected number of clusters with radius $r$, denoted here as $N_r$, depends on the free energy difference between the solid and liquid phases $\Delta G_r$,

\begin{equation*}
    N_r = N_0 \exp \left(
        - \frac{\Delta G_r}{k_B T}
    \right),
\end{equation*}

\noindent with $N_0$ a constant, $T$ the temperature, and $k_B$ the Boltzmann constant.
The free energy difference assumes spherical clusters and balances the volumetric free energy difference between the solid-liquid phases $\Delta g_\mathrm{SL}$ and the interfacial free energy $\gamma_\mathrm{SL}$,

\begin{equation*}
    \Delta G_r =
        \frac{4}{3} \pi r^3 \Delta g_\mathrm{SL} +
        4 \pi r^2 \gamma_\mathrm{SL}.
\end{equation*}

\begin{figure}[!h]
    \centering
    \includegraphics[width=0.7\linewidth]{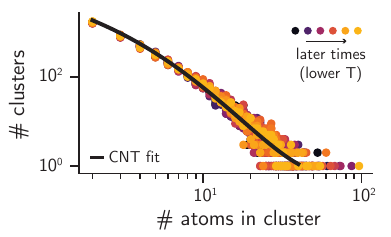}
    \caption{
    Cluster size distribution in the melt prior to nucleation. The size distribution follows a power law similar to predictions from the CNT (fitted black line).
    }
    \label{fig:si:04-solid-cna}
\end{figure}

\noindent The fit in Fig. \ref{fig:si:04-solid-cna} is obtained by fitting the unknowns $N_0$, $\Delta g_\mathrm{SL}$, and $\gamma_\mathrm{SL}$ for the equation

\begin{equation*}
    \log N_r = \log N_0 -
        \frac{4 \pi r^3}{3 k_B T} \Delta g_\mathrm{SL} -
        \frac{4 \pi r^2}{k_B T} \gamma_\mathrm{SL}.
\end{equation*}

\noindent In this case, the values of $r$ are estimated from the cluster size from the graph-theoretical approach and a density of $8960$ kg/m$^3$.
The fit was performed for the temperature of $917$ K, which is approximately the temperature of solidification during the simulation, and used all cluster sizes of the first 120 steps of the simulation.
The nucleation event is observed at the 125th step.

\subsection*{Uncertainty quantification}
\label{sec:si:methods-uq}

\begin{figure}[!ht]
    \centering
    \includegraphics[width=0.8\linewidth]{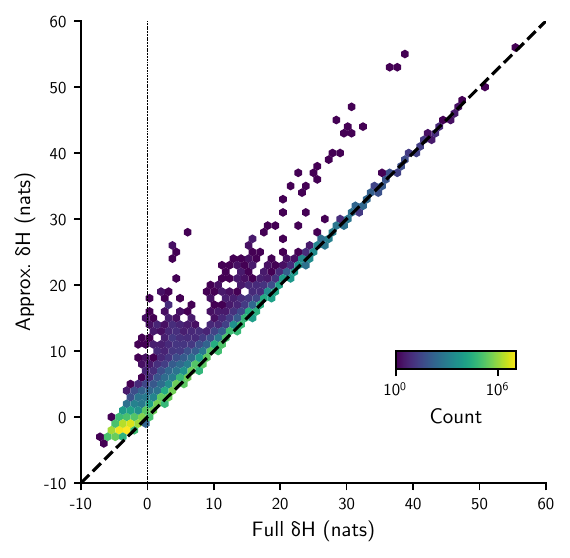}
    \caption{
    Comparison between approximated value of $\dH$ computed with a number of neighbors of $n=30$ queried per data point and with $m=50$ neighbors when constructing the index. The comparison shows that, in most cases, the use of approximate $\dH$ is a reasonable approximation. Importantly, due to the results in Eq. \eqref{eq:si:dh-kde-knn}, the approximation always overestimates the value of $\dH$, and thus can serve as an upper bound for the differential entropy.
    }
    \label{fig:si:04-approx-dH}
\end{figure}

\noindent\textbf{Approximate nearest neighbors:}
The approximate nearest neighbors for feature vectors $\X$ demonstrated in Sec. \ref{sec:si:approx-dH} was computed using PyNNDescent (v. 0.5.11), that implements a search strategy based on $k$-neighbor graph construction.\citeMain{dong2011efficient}
The number of neighbors used to construct the index is represented with $m$ in Fig. \ref{fig:si:04-approx-dH}.
The default number of trees, leaf sizes, and other parameters were used in the construction of the index.
Searches were performed using an epsilon value of 0.1.

\clearpage
\section{Supplementary Figures}

\begin{figure}[!h]
    \centering
    \includegraphics[width=\linewidth]{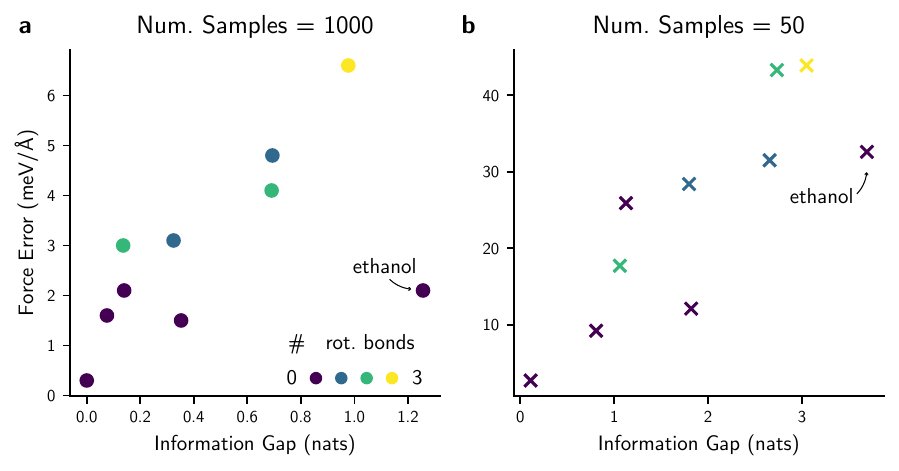}
    \caption{
    Correlation between the force errors from a MACE model and the information gap for each molecule.
    The errors are shown separately for the model trained on two dataset sizes: \textbf{a}, 1000 samples and \textbf{b}, 50 samples.
    The color represents the number of rotatable bonds for each molecule.
    Ethanol is an outlier from the trend in \textbf{a}.
    }
    \label{fig:si:04-rmd17-const}
\end{figure}

\begin{figure}[!h]
    \centering
    \includegraphics[width=0.5\linewidth]{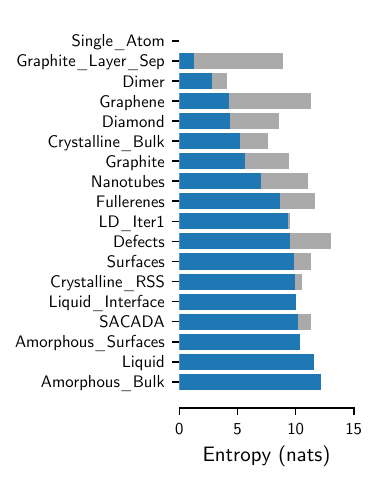}
    \caption{
    Entropies for the subsets of GAP-20 for carbon (blue) and maximum possible entropy for each subset ($\log n$, gray).
    The entropy of the ``Single\_Atom'' subset is zero, as it contains only a single environment.
    }
    \label{fig:si:04-gap20-entropy}
\end{figure}

\begin{figure}[!h]
    \centering
    \includegraphics[width=\linewidth]{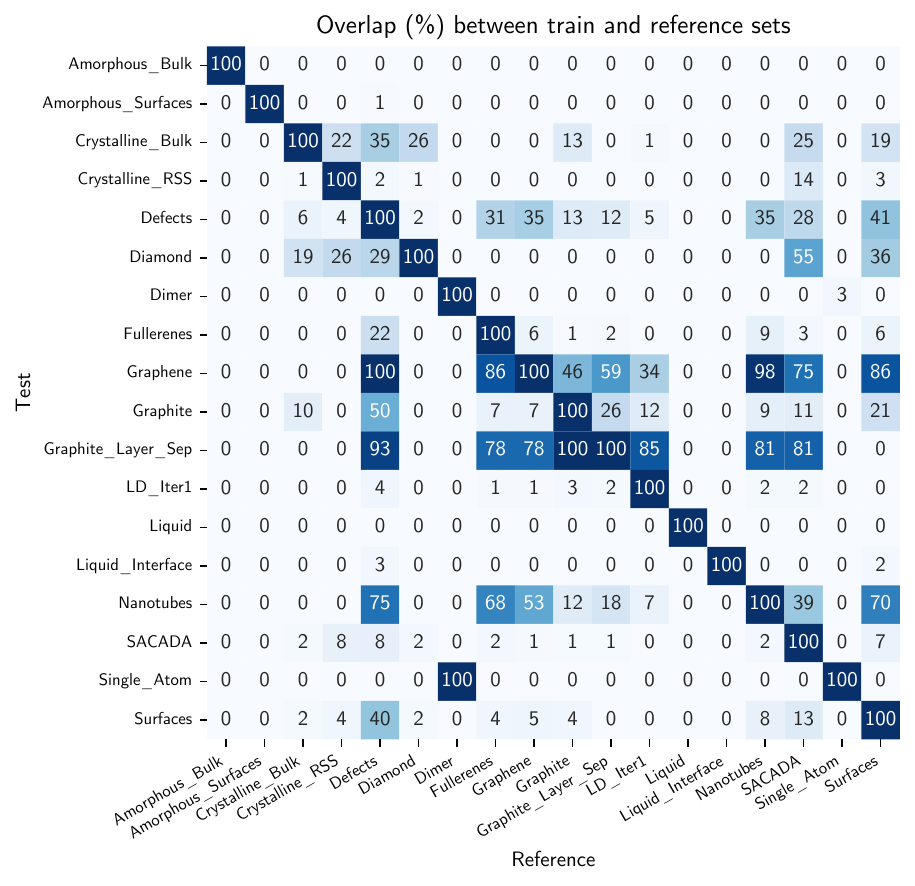}
    \caption{
    Overlap between test and reference sets for the GAP-20 carbon dataset.
    }
    \label{fig:si:04-gap20-dH-table}
\end{figure}

\begin{figure}[!h]
    \centering
    \includegraphics[width=\linewidth]{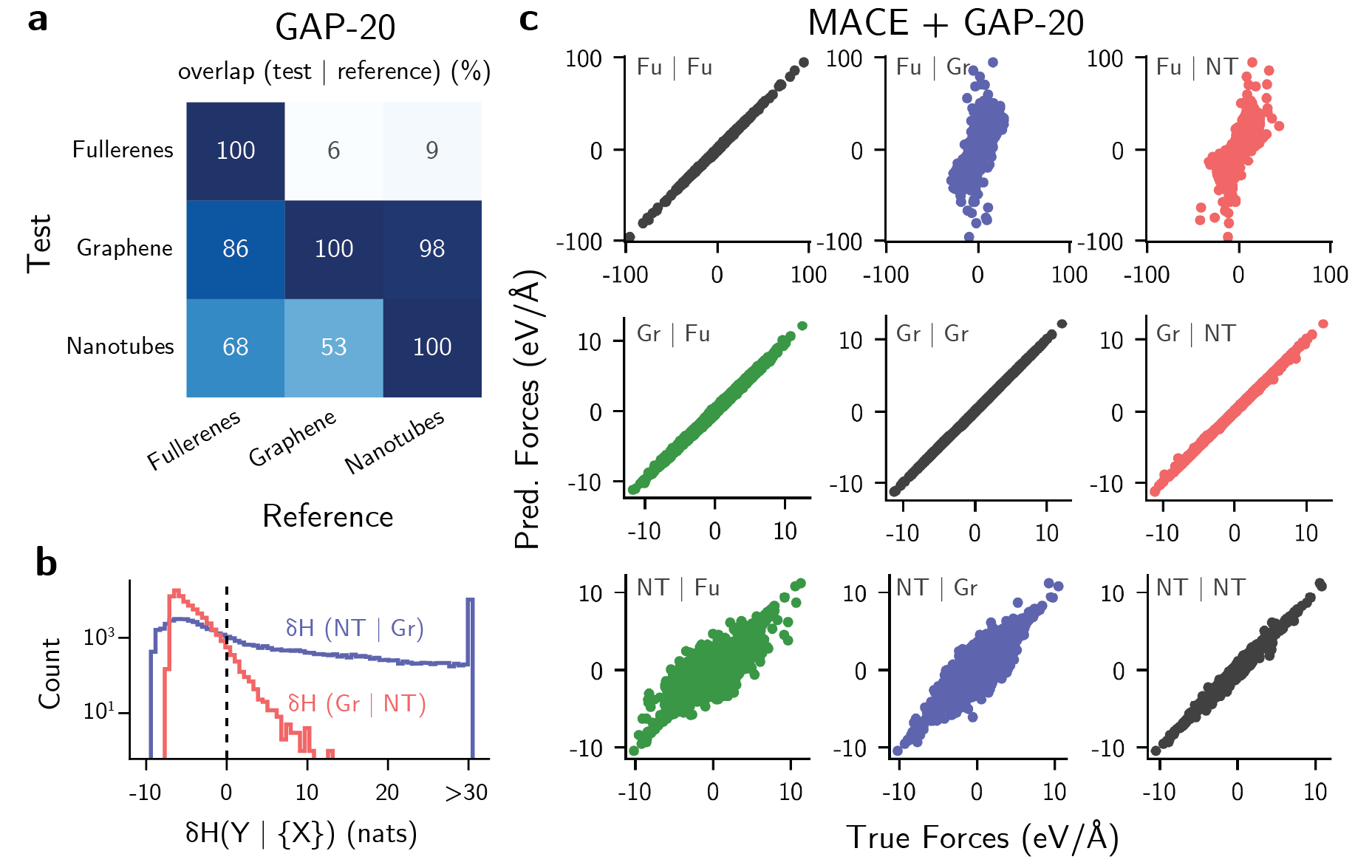}
    \caption{
    \textbf{a}, Overlaps between different subsets of the GAP-20 dataset, as also shown in Fig. \ref{fig:03-uq}a of the main text.
    The overlap is computed from the distribution of $\dH$ values such as the one in \textbf{b} that illustrates the asymmetry between the ``Graphene'' and ``Nanotubes'' datasets.
    \textbf{c}, parity plot illustrating the prediction errors of a MACE model trained on the ``Fullerenes'' (Fu), ``Graphene'' (Gr), and ``Nanotubes'' (NT) subsets, and tested on the others.
    The results reflect the trends of overlap in \textbf{a}.
    }
    \label{fig:si:03-uq-parity}
\end{figure}

\begin{figure}[!h]
    \centering
    \includegraphics[width=0.8\linewidth]{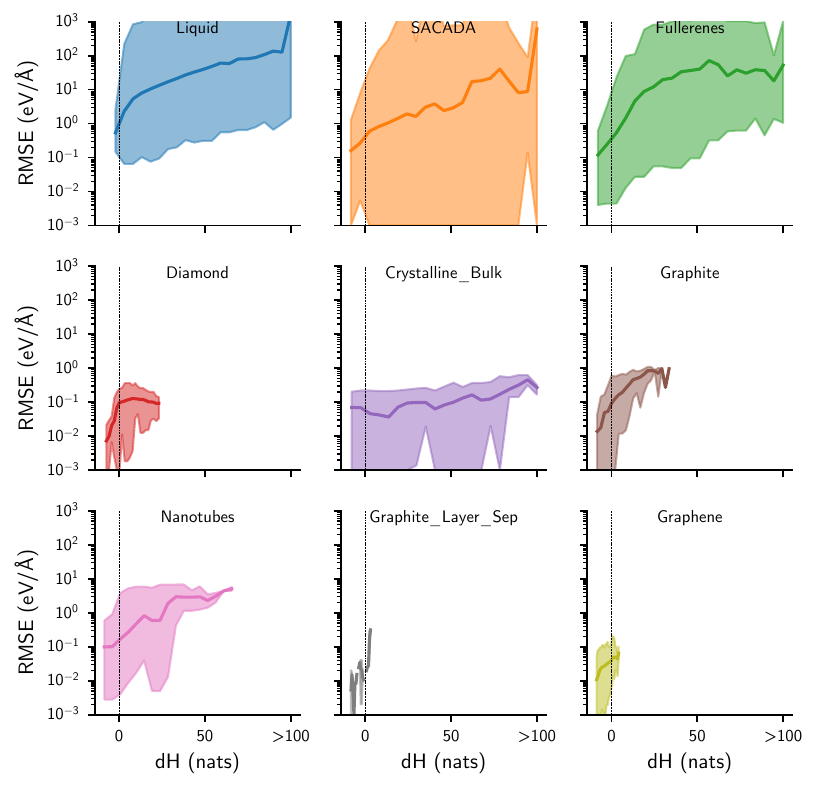}
    \caption{
    RMSE of forces for MACE models trained on the ``Defects'' GAP-20 dataset and tested on other subsets.
    The test splits are sorted by increasing overlap with the training sets.
    The shaded area represents the range of the error distribution in each window of $\dH$.
    For clarity, small errors are truncated to be equal to 1 meV/\AA, and the plot is truncated at a maximum error of 1000 eV/\AA.
    Because some data points in the ``Liquid'' or ``SACADA'' subsets are infinitely far away from the ``Defects'' training set, their values of $\dH$ are also infinite.
    To avoid issues with the visualization, we clipped the values of $\dH$ at 100 for all sets.
    }
    \label{fig:si:03-uq-defects-dH}
\end{figure}

\begin{figure}[!h]
    \centering
    \includegraphics[width=0.7\linewidth]{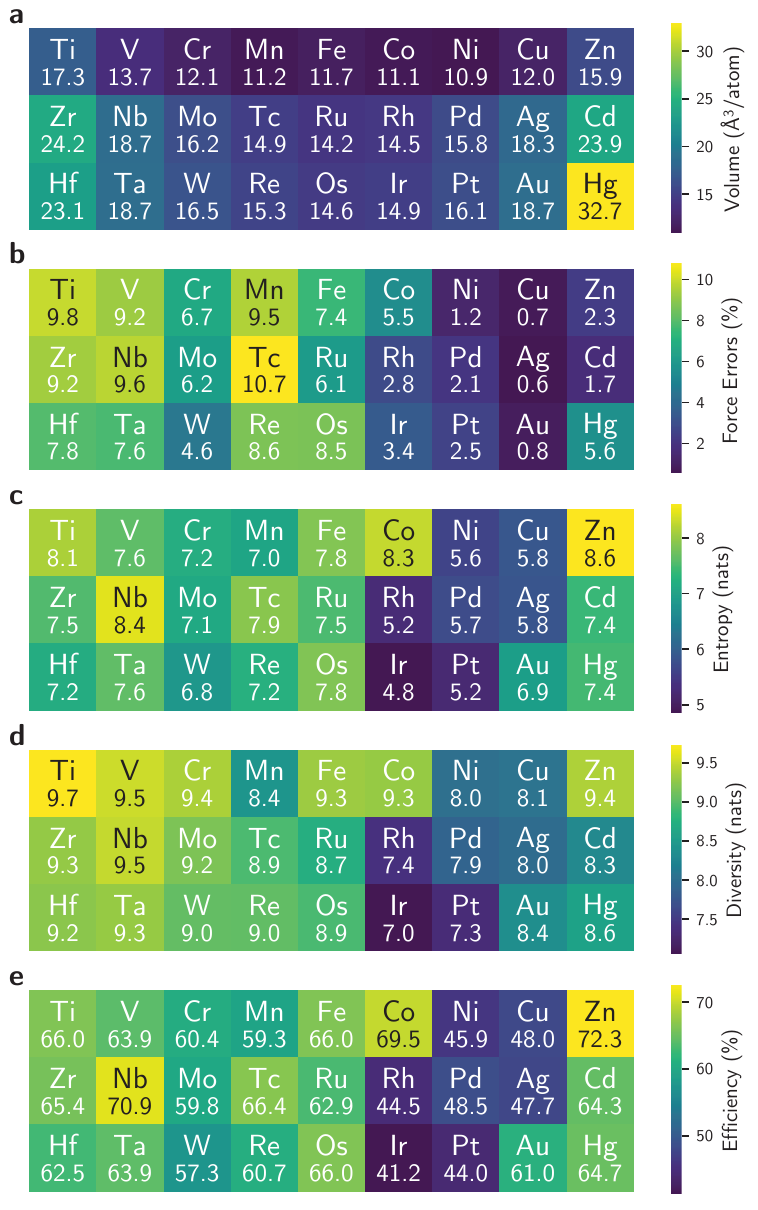}
    \caption{
        Per-element values for the full TM23 (cold, warm, and melt subsets) for: \textbf{a}, atomic volume, \textbf{b}, relative force errors for the NequIP models reported by Owen \textit{et al.}, \textbf{c}, entropy, \textbf{d}, diversity, and \textbf{e}, efficiency of each dataset.
        The efficiency is defined as the value of entropy divided by the logarithm of the number of environments ($\log n$).
    }
    \label{fig:si:05-tm23-tables}
\end{figure}

\begin{figure}[!h]
    \centering
    \includegraphics[width=\linewidth]{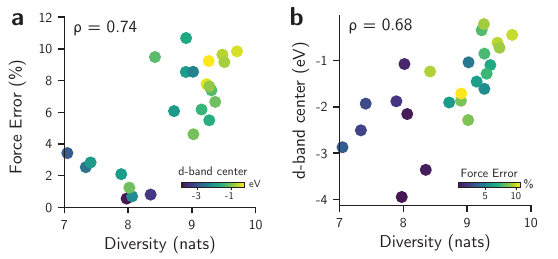}
    \caption{
        Correlation between \textbf{a}, Force errors from Owen \textit{et al.} and diversity, and \textbf{b}, d-band center of the transition metals, as reported by Owen \textit{et al.}, and dataset diversity.
        The Pearson correlation coefficients are also given.
        The results are described without group 12 elements, following the discussion by Owen \textit{et al.}
    }
    \label{fig:si:05-tm23-diversity}
\end{figure}

\begin{figure}[!h]
    \centering
    \includegraphics[width=\linewidth]{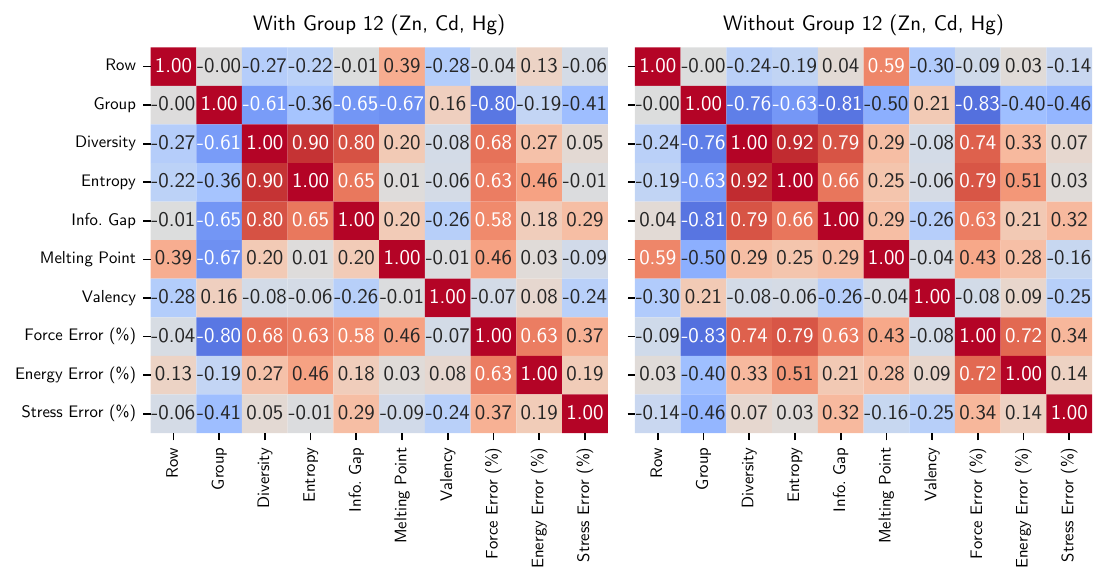}
    \caption{
        Pearson correlation coefficients between quantities describing the TM23 dataset.
        The results are described with (left) and without (right) group 12 elements, following the discussion by Owen \textit{et al.}
        In addition to the known trend between force errors and the group of the periodic table, we demonstrate that the diversity, entropy, and information gap also exhibit a reasonably strong correlation with the force errors.
    }
    \label{fig:si:05-tm23-corr}
\end{figure}

\begin{figure}[!h]
    \centering
    \includegraphics[width=\linewidth]{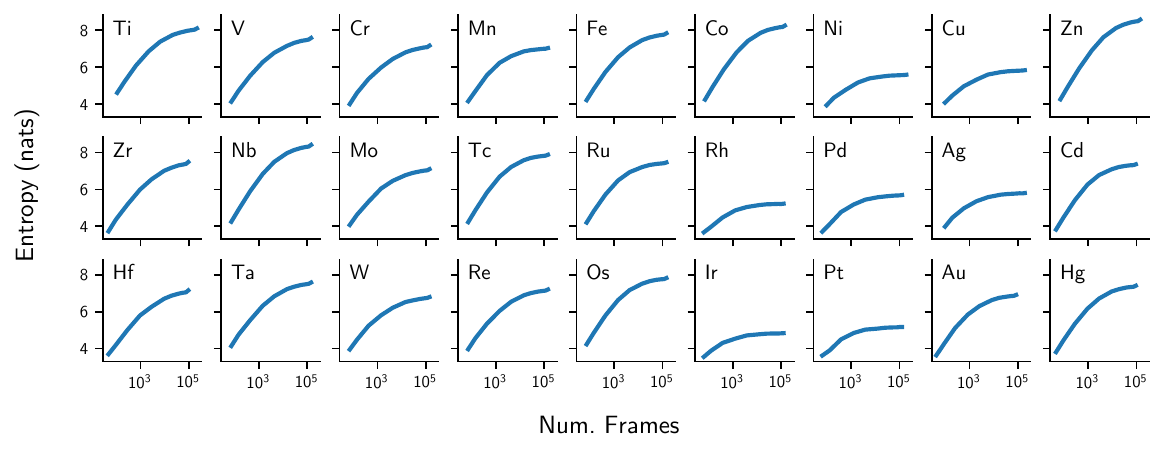}
    \caption{
        Information  (learning) curves for the TM23 dataset obtained for the full subsets per element.
        The entropy is computed by averaging five different runs to obtain reliable statistics on each subset.
        Subsets are obtained by randomly sampling the number of environments from the main dataset.
    }
    \label{fig:si:05-tm23-learning}
\end{figure}

\begin{figure}[!h]
    \centering
    \includegraphics[width=\linewidth]{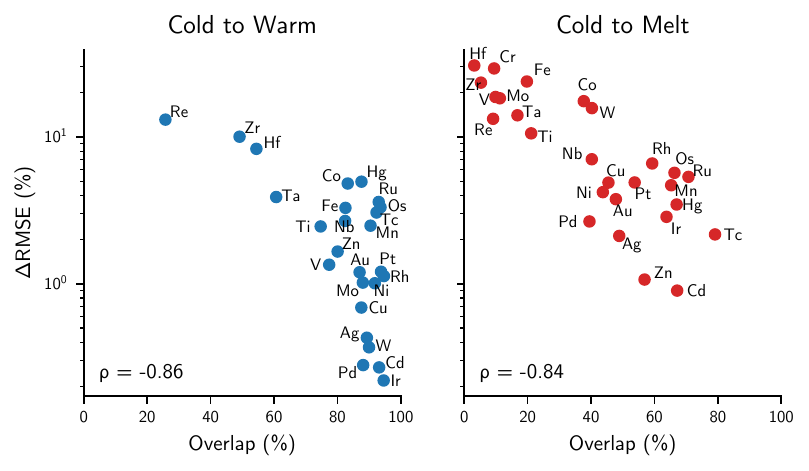}
    \caption{
        Per-element overlap between train and test sets in the transferability experiments of TM23, and impact in the relative error increase. The left chart shows the per-element results for models tested on the ``warm'' subset, and the right chart shows the per-element results for models tested on the ``melt'' subset. Errors are obtained as reported by Owen \textit{et al.} The correlation coefficients are similar for both systems, suggesting they follow reasonably similar power laws. This figure is identical to Fig. \ref{fig:05-tm23}f in the main text.
    }
    \label{fig:si:05-tm23-overlaps}
\end{figure}

\begin{figure}[!h]
    \centering
    \includegraphics[width=\linewidth]{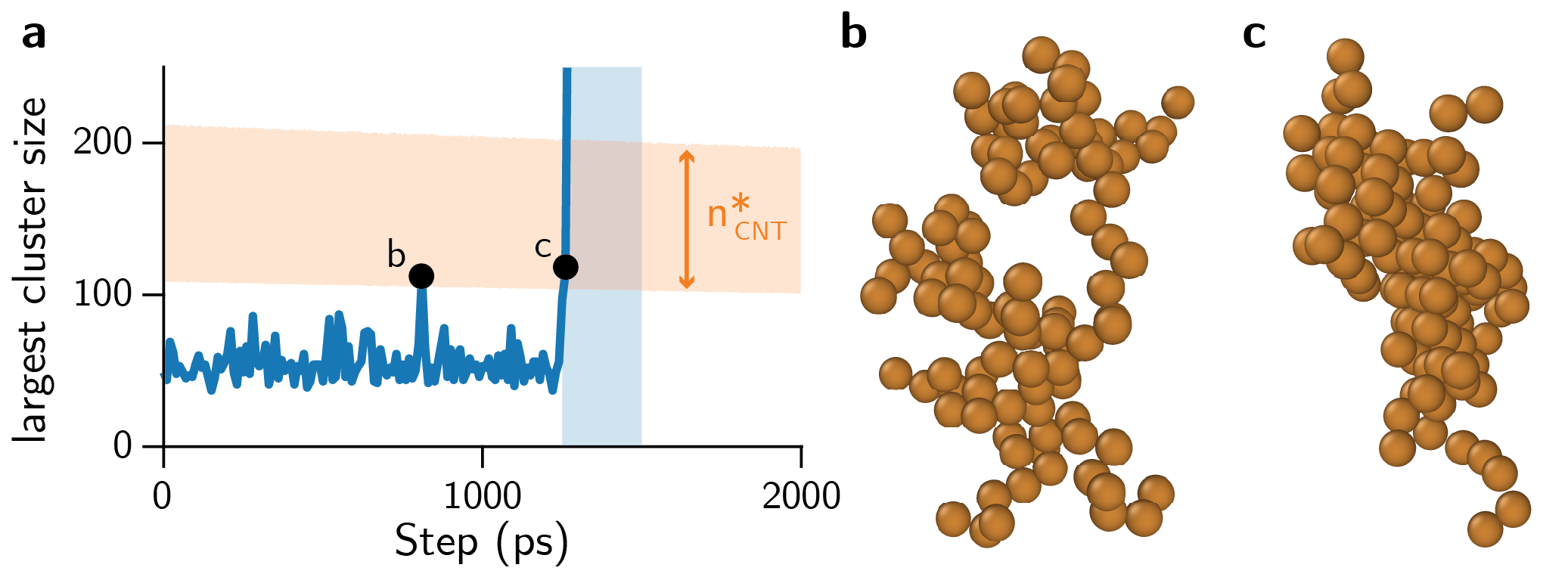}
    \caption{
        \textbf{a}, maximum cluster size throughout the solidification trajectory, as depicted in Fig. \ref{fig:04-outlier}f of the main paper.
        The black dots indicate two frames when the maximum cluster size surpasses the minimum required for nucleation.
        The visualization of these two clusters is shown in \textbf{b} and \textbf{c}.
        Whereas both have approximately the same number of atoms, \textbf{b} is much less compact compared to \textbf{c}, and may be better represented by two separate clusters instead of one.
        This may be an artifact of the graph-theoretical approach used to identify connected atoms in the simulation cell given values of $\dH$.
    }
    \label{fig:si:04-nuclei}
\end{figure}

\begin{figure}[!h]
    \centering
    \includegraphics[width=\linewidth]{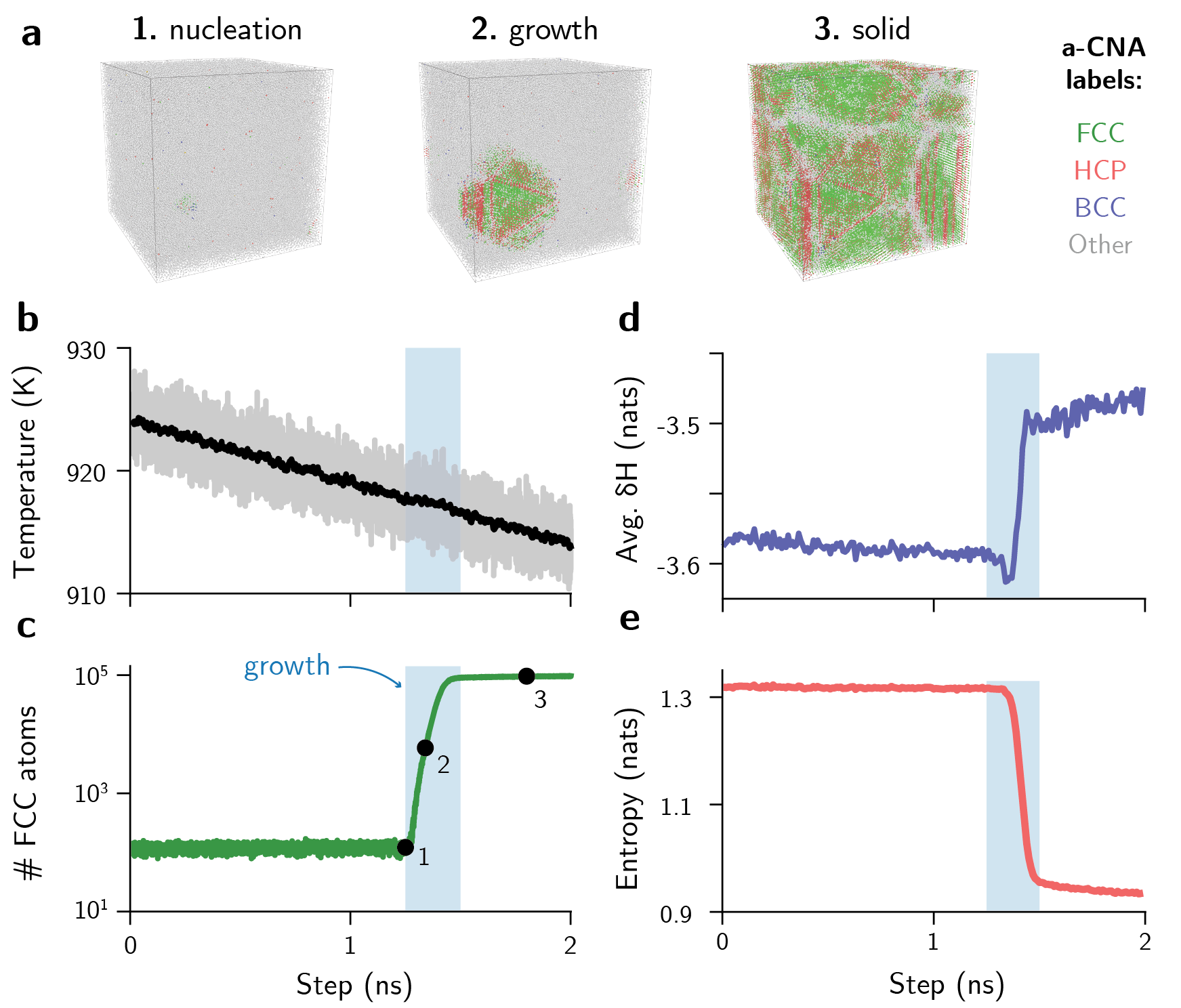}
    \caption{
    \textbf{a}, Visualization of the solidification trajectory during the nucleation, growth, and solid states.
    FCC, HCP, and BCC phases are shown with green, red, and blue colors, respectively.
    Phases not identified by a-CNA are represented in gray.
    \textbf{b}, Average (black) and instantaneous (gray) temperature and \textbf{c}, number of FCC atoms derived from the MD simulation.
    The shaded blue area indicates the time window where crystal growth is observed.
    The critical nucleus is observed around 917 K.
    The black dots indicate the frames corresponding to nucleation, growth, and final solidified system visualized in \textbf{a}.
    \textbf{e}, Average $\dH$ using the first frame (melt) as reference for the entire solidification trajectory.
    The drop in the average $\dH$ around 1.25 ns suggests that the phases during growth are well-represented in the melt.
    \textbf{e}, Entropy computed for each frame using our information theoretical method.
    }
    \label{fig:si:04-solid-curves}
\end{figure}

\begin{figure}[!ht]
    \centering
    \includegraphics[width=0.8\linewidth]{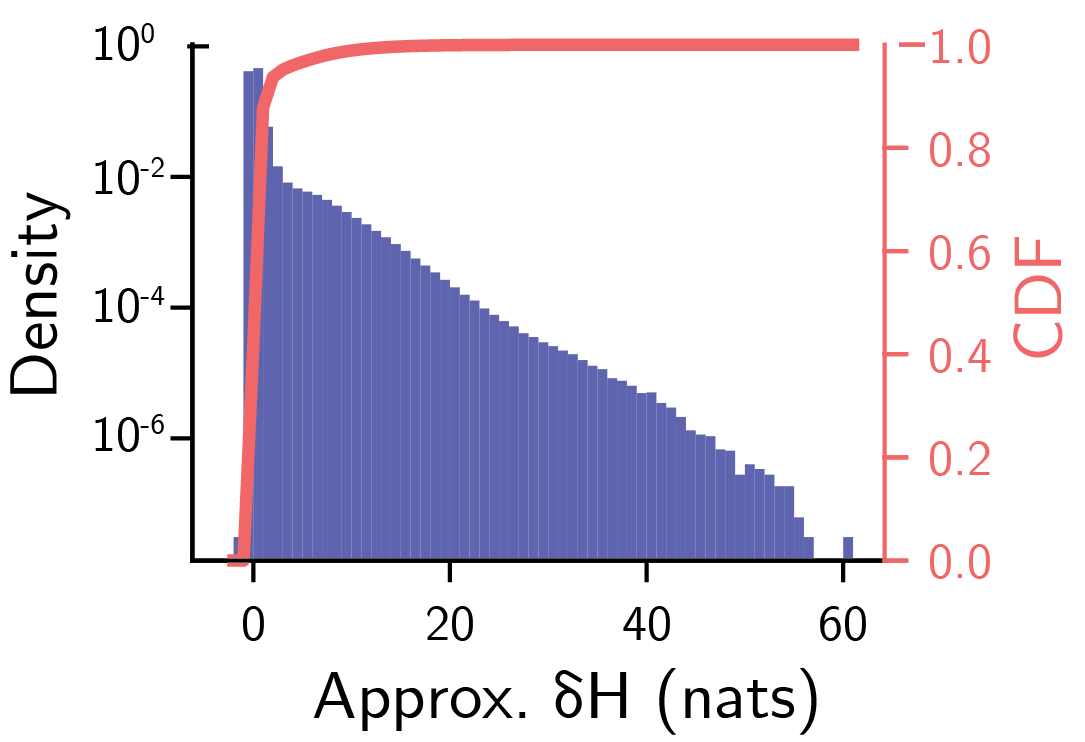}
    \caption{
    Distribution (blue) and cumulative density function (CDF, red) of estimated $\dH$ values.
    87\% of the atoms exhibit $\dH < 0$ nats and thus are reasonably close to the training set.
    }
    \label{fig:si:05-Ta-density}
\end{figure}

\clearpage
\section{Supplementary Tables}

\begin{table}[!htb]
\caption{Overlap between test and train sets for the TM23 dataset, as computed by the fraction of differential entropy values larger than 0 ($\dH > 0$). The overlaps are provided in \%.}
\label{tab:si:05-tm23-overlaps}
\centering
\begin{tabular}{cccc}
\toprule
Element & Full Dataset & Cold → Warm & Cold → Melt \\
\midrule
Ag & 99.6 & 89.3 & 48.9 \\
Au & 98.4 & 87.0 & 47.8 \\
Cd & 98.0 & 93.2 & 67.2 \\
Co & 97.0 & 83.3 & 37.7 \\
Cr & 95.4 & 87.6 & 9.4 \\
Cu & 99.5 & 87.5 & 45.5 \\
Fe & 95.8 & 82.5 & 19.8 \\
Hf & 94.8 & 54.5 & 3.2 \\
Hg & 97.3 & 87.6 & 67.0 \\
Ir & 99.8 & 94.6 & 63.8 \\
Mn & 98.9 & 90.5 & 65.3 \\
Mo & 96.6 & 88.0 & 11.2 \\
Nb & 95.3 & 82.4 & 40.2 \\
Ni & 99.4 & 91.8 & 43.7 \\
Os & 98.5 & 93.6 & 66.3 \\
Pd & 99.3 & 88.1 & 39.5 \\
Pt & 99.5 & 93.8 & 53.8 \\
Re & 97.9 & 25.8 & 9.1 \\
Rh & 99.5 & 94.7 & 59.3 \\
Ru & 98.6 & 93.0 & 70.7 \\
Ta & 96.0 & 60.7 & 16.8 \\
Tc & 98.5 & 92.3 & 79.1 \\
Ti & 95.1 & 74.7 & 21.1 \\
V & 94.8 & 77.4 & 9.9 \\
W & 99.0 & 90.0 & 40.3 \\
Zn & 97.2 & 80.1 & 56.9 \\
Zr & 93.5 & 49.1 & 5.3 \\
\bottomrule
\end{tabular}
\end{table}


\end{document}